\documentclass[10pt]{article}
\usepackage[english]{babel}
\usepackage{sansmathfonts}
\usepackage[T1]{fontenc}
\usepackage{amsmath,amsfonts,amssymb,amstext,amsmath,amsthm,amscd}
\usepackage{mathtools}
\usepackage[a4paper,top=2cm,bottom=2.5cm,left=2.5cm,right=2.5cm]{geometry}
\usepackage{breakcites}
\usepackage{mathtools}
\usepackage{authblk}

\usepackage{enumitem}
\usepackage{booktabs}
\usepackage{mathtools}
\usepackage{mathrsfs}
\usepackage{dsfont}
\usepackage{enumitem}
\usepackage{graphicx}
\usepackage{subcaption}
\usepackage{wrapfig}
\usepackage{indentfirst}
\usepackage{latexsym}
\usepackage{sidecap}
\usepackage{booktabs}
\usepackage{hyperref}

\usepackage{lipsum}
\usepackage{cleveref}
\usepackage[dvipsnames]{xcolor}
\usepackage[nodisplayskipstretch]{setspace}

\newcommand\myshade{85}
\colorlet{mylinkcolor}{violet}
\colorlet{mycitecolor}{YellowOrange}
\colorlet{myurlcolor}{RoyalBlue}

\hypersetup{
  colorlinks=true,
  linkcolor=myurlcolor,
  citecolor=mycitecolor!\myshade!black,
  filecolor=magenta,      
  urlcolor=magenta,
}
\usepackage{setspace}
\usepackage{breakcites}
\usepackage{mathrsfs}
\usepackage{textcomp}
\usepackage{braket}
\usepackage{colortbl}
\usepackage{bbm}
\usepackage{comment}
\usepackage[colorinlistoftodos]{todonotes}\usepackage{mathrsfs}
\usepackage{textcomp}
\usepackage{braket}
\usepackage{colortbl}
\usepackage{bbm}
\usepackage{comment}
\usepackage[colorinlistoftodos]{todonotes}
    
\newcommand{\yf}{$\bar{y}_F$}

 \usepackage{lineno}

\newcommand{\dt}{D_{10}}

\usepackage{tikz}
\usetikzlibrary{arrows}

\title{One scale to rule them all: interpretable multi-scale Deep Learning for predicting cell survival after proton and carbon ion irradiation}

\author[1,2]{Giulio Bordieri}
\author[3]{Giorgio Cartechini}
\author[4]{Anna Bianchi}
\author[4]{Anna Selva}
\author[4]{Valeria Conte}
\author[2,5,$\dag$]{Marta Missiaggia}
\author[2,6,$\dag$,*]{Francesco G. Cordoni}

\affil[1]{Department of Physics, University of Trento, via Sommarive 14, Trento, 38123, Italy}
\affil[2]{Trento Institute for Fundamental Physics and Application (TIFPA), via Sommarive 15, Trento, 38123, Italy}
\affil[3]{Department of Radiation Oncology (Maastro), GROW School for Oncology and Reproduction, Maastricht University Medical Centre, Maastricht, The Netherlands}
\affil[4]{Istituto Nazionale di Fisica Nucleare - Laboratori Nazionali di Legnaro, Viale dell’Università 2, 35020 Legnaro PD, Italy}
\affil[5]{Department of Physics \& Astronomy, Louisiana State University (LSU), 202 Nicholson Hall, Baton Rouge, LA 70803, US}
\affil[6]{Department of Civil, Environmental and Mechanical Engineering, University of Trento, via Mesiano 77, Trento, 38123, Italy}
\affil[$\dag$]{Equal contribution}
\affil[*]{corresponding author: francesco.cordoni@unitn.it}

\begin{document}

\maketitle   

\begin{abstract}
The relationship between the physical characteristics of the radiation field and biological damage is central to both radiotherapy and radioprotection, yet the link between spatial scales of energy deposition and biological effects remains not entirely understood. 
To address this, we developed an interpretable deep learning model that predicts cell survival after proton and carbon ion irradiation, leveraging sequential attention to highlight relevant features and provide insight into the contribution of different energy deposition scales. Trained and tested on the PIDE dataset, our model incorporates, beside LET,  nanodosimetric and microdosimetric quantities simulated with MC-Startrack and Open-TOPAS, enabling multi-scale characterization. 
While achieving high predictive accuracy, our approach also emphasizes transparency in decision-making. We demonstrate high accuracy in predicting RBE for in vitro experiments. Multiple scales are utilized concurrently, with no single spatial scale being predominant. Quantities defined at smaller spatial domains generally have a greater influence, whereas the LET plays a lesser role.  
\end{abstract}

\tableofcontents

\section{Introduction}

Ionizing radiation therapy remains a cornerstone in the treatment of approximately half of patients with localized malignant tumors \cite{baskar2012cancer, durante2019charged}. Among the available modalities, particle therapy, particularly with protons and carbon ions, has gained clinical adoption due to its superior dose distribution and enhanced tissue sparing compared to conventional X-rays \cite{durante2010charged}. By the end of 2023, more than 410,000 patients worldwide had received particle therapy, with nearly 350,000 treated using protons \cite{PTCOG}.

As particle therapy continues to expand, a deeper understanding of the biological effects of radiation is essential for improving therapeutic efficacy while minimizing adverse outcomes. Radiobiological models must therefore account for both the inherently stochastic nature of radiation-matter interactions and the complex downstream biological response mechanisms.

Recent research has increasingly focused on optimizing physical parameters to influence biological outcomes, with particular emphasis on linear energy transfer (LET) optimization, microdosimetric approaches, and nanodosimetric approaches \cite{romano2014monte, faddegon2023ionization, cartechini2024integrating}. These strategies differ primarily in the spatial scale at which energy deposition is considered. LET optimization addresses average energy deposition along particle tracks at a macroscopic level. Microdosimetry examines energy deposition patterns at the micrometer scale, relevant to cellular structures. Nanodosimetry focuses on the nanometer scale, where individual ionization clusters within DNA-sized volumes are analyzed.

However, these strategies often neglect the underlying complexity of radiation stochasticity at different scales, focusing instead on a single domain size over which energy deposition is considered. One open question remains as to whether a specific domain size exists that is most relevant for predicting the biological effect of radiation and, consequently, treatment efficacy. In fact, importantly, it is believed these three approaches are complementary, as different spatial scales are associated with distinct biological endpoints: macroscopic approaches often relate to tumor control and normal tissue complication probabilities, microdosimetric analyses inform cell survival and repair processes, while nanodosimetric studies are linked to complex DNA damage and mutagenesis \cite{missiaggia2024radiation}. In this context, the introduction of both nano- and micro-scale perspectives alongside the commonly used macroscopic LET appears essential. Different scales likely contribute complementary information, as each reflects energy deposition in different types of biological targets \cite{friedrich2018dna}. While it is widely acknowledged that radiation effects manifest across multiple spatial scales, there is currently no consensus on which of these provides the most accurate or mechanistically informative link to cellular survival outcomes \cite{missiaggia2024radiation}.

This work provides a detailed data-driven investigation of radiation quality across multiple spatial domains by integrating nanodosimetric and microdosimetric information alongside conventional dosimetric variables. It addresses the complex challenge of understanding how energy deposition described at different levels affects the overall biological impact of radiation in terms of cell survival.
In this context, Deep Learning (DL) methods allow us to identify patterns and make predictions directly from data. DL relies on artificial neural networks composed of multiple layers that can automatically extract and combine features from large datasets to capture nonlinear dependencies and complex interactions among features. Despite the great predictive accuracy of DL, a major challenge lies in its interpretability. 
To address this issue, we develop a DL model to predict the biological effects of radiation on tissues, achieving both high predictive accuracy and built-in interpretability through an attention mechanism. We leverage sequential attention to dynamically select relevant features at each decision step, enabling native interpretability and efficient learning through sparse attention, forcing the model to focus on specific features at each step. This approach allows us to gain insights into the contribution of different energy deposition scales to biological damage, while maintaining transparency in the model’s decision-making process.

A DL model for cell survival prediction is trained to learn patterns from radiobiological and dosimetric features and evaluated on independent datasets to assess its ability to generalize and provide biologically interpretable insights into post-irradiation outcomes. In this work, we focus on proton and carbon ion cell survival experiments contained in the Particle Irradiation Data Ensemble (PIDE) dataset \cite{friedrich2021update}. To enrich the physical description beyond conventional macroscopic parameters, we augment the dataset by simulating microdosimetric spectra using the open-source TOPAS toolkit \cite{perl2012topas, faddegon2020topas}, and nanodosimetric distributions following the approach described in \cite{grosswendt2014upgraded}.
The importance of considering multiple spatial scales in radiobiological modeling has also been recently highlighted in \cite{friedrich2018dna}, where the authors conducted a model-based investigation of how different scales are incorporated into mathematical formulations. Our goal is conceptually similar, that is, understanding the interplay between scales, but we approach it from a fundamentally different perspective. Instead of imposing assumptions through predefined mathematical structures, we develop a DL model that is inherently transparent, allowing interpretation of its decision-making process. This design ensures that no bias is introduced in the model construction; rather, the contribution of different scales emerges directly from the data.

We train the model to predict cell survival following irradiation with protons and carbon ions, which represents the most widely studied biological endpoint in radiotherapy and particle therapy, supported by decades of experimental evidence. Cell survival is particularly relevant because it reflects the cumulative effect of processes occurring across multiple scales—from DNA damage induction at the nanometer level, through cellular repair dynamics, to population-level outcomes. By considering different particle types and clinically relevant energy ranges, our approach enables investigation of how physical and biological scales interact under varying irradiation conditions, providing new insights into the multiscale nature of radiation response.

It is important to emphasize that our primary aim is not to introduce a more accurate predictive model, but rather to investigate interpretability in multiscale radiobiological modeling. Although our DL framework achieves state-of-the-art performance in predicting cell survival, the focus is on exploring how the model reaches its decisions and discerning the relative contributions of different spatial scales of energy deposition. Whereas conventional mathematical models impose scale dependencies through their structure, our transparent DL model allows such relationships to emerge organically from the data—free from structural bias. This enables a clear analysis of nanodosimetric, microdosimetric, and macroscopic descriptors and how they jointly affect cell survival.

In addition to modeling survival curves, we concentrate on three biologically and clinically significant endpoints. First, the dose to 10\% survival ($D_{10}$), historically the benchmark for calculating relative biological effectiveness (RBE) in particle therapy \cite{ICRP92}. RBE is formally defined as
\begin{equation}\label{EQN:RBE}
\text{RBE} = \frac{D_{\mathrm{ref}}}{D_{\mathrm{reference}}}\Big|_{\,\text{same biological effect}}
\end{equation}
where $D_{\mathrm{ref}}$ is the dose of reference radiation (e.g., low-LET X-rays), and $D_{\mathrm{test}}$ is the dose of the test radiation (e.g., protons or carbon ions) required to achieve the same biological effect, typically 10\% clonogenic survival. RBE remains a cornerstone in treatment planning since it quantifies how much more (or less) effective a radiation quality is relative to photons, directly guiding dose prescription and clinical outcomes.

Second, we consider the initial slope of the survival curve (the $\alpha$-region), reflecting intrinsic radiosensitivity and repair mechanisms at low doses. This endpoint has long been central to radiobiology, especially in evaluating normal tissue response.

Finally, we evaluate survival at 2~Gy, which aligns with the conventional fractional dose in radiotherapy, providing a standard benchmark for both modality comparison and biological modeling.

By combining predictive accuracy, interpretability, and clinically meaningful endpoints, our framework offers a novel lens through which to understand how different scales of energy deposition impact biological outcomes. This not only validates model performance but also advances mechanistic insights essential for optimizing particle therapy and personalizing treatment protocols.

\section{Materials and Methods}

\subsection{PIDE dataset}

The model was trained and validated using data from the Particle Irradiation Data Ensemble (PIDE, version 3.2) \cite{friedrich2021update}, which includes cell survival experiments for both protons and carbon ions. The PIDE provides detailed specifications of the radiation fields used in these experiments. Leveraging this information, we reproduced the experimental conditions through Monte Carlo (MC) simulations to extract additional physical descriptors, including nanodosimetric and microdosimetric quantities.

Cell survival following irradiation was modeled using the linear-quadratic (LQ) formalism, which expresses survival as
\[
S(D) = \exp(-\alpha D - \beta D^2),
\]
where $D$ is the absorbed dose, and $\alpha$ and $\beta$ are two parameters that depend on both the biological system and the radiation field. In the following, we denote by $\alpha_X$ and $\beta_X$ the parameters corresponding to the reference radiation. The ratio $\alpha/\beta$ denotes the dose at which these two components are equal and is widely used to characterize tissue radiosensitivity. This parameter is fundamental in radiotherapy for guiding fractionation schemes and comparing biological effects across radiation qualities. Following previous works \cite{cordoni2023artificial, pfuhl2022comprehensive}, the DL model was trained on parameters derived from an exponential LQ fit of the survival curves rather than raw survival fractions. This choice addresses the substantial variability observed in reported survival data across experiments, which can introduce noise and bias. To ensure robust curve fitting, experiments reporting fewer than three measurement points were excluded, as at least three data points are required to estimate LQ parameters reliably. 

The dataset was split into three folds at the level of experiments, ensuring that all data from a single experiment were assigned to the same fold. This strategy prevents data leakage across folds and allows for a more robust evaluation of generalization. The model was trained iteratively on two folds and validated on the remaining fold, cycling through all three combinations. This approach enables out-of-sample validation across the entire dataset, which is particularly important given the limited amount of available data. A standard train–test split would not provide sufficient robustness for the test set under these constraints. All results reported in this work refer to performance on the three validation sets. 

Logarithmic transformation of the surviving fraction data is applied to smooth training. Further, the Box-Cox transformation of LET and microdosimetric values is applied to stabilize variance and improve normality.

%The model is trained using 5-fold cross-validation with batch size 64, validation split 20\%, and learning rate 0.002 for 100 epochs. The cost function is based on RMSE and optimized using Adam \cite{zhang2018improved}. Post-processing applies a linear-quadratic model fit to predicted results.

For this study, 320 different experiments are selected, covering the full available PIDE cohort of irradiations with protons and carbon ions. This choice allows us to cover the LET, respectively. beam energy, range from 1 keV/$\mu$m to 576 keV/$\mu$m, resp. 0.7 MeV/u to 440 MeV/u. 

In the following, we will compare the developed DL model with two standard models used in the clinics, the Microdosimetric Kinetic Model (MKM) \cite{inaniwa2018adaptation} and the Local Effect Model (LEM) version III \cite{elsasser2008accuracy}. As done in previous work \cite{cordoni2023artificial}, the model's implementation and parameters are reported in \cite{manganaro2018survival}. The comparison is performed on the HSG and V79 cell lines being the ones where the LEM and MKM have been extensively calibrated and validated.

\subsection{Nanodosimetry}

Nanodosimetry seeks to characterize the stochastic nature of energy deposition at nanometer scales by analyzing ionization cluster‑size distributions (ICSD) within target volumes comparable to segments of cellular DNA \cite{rabus2011nanodosimetry, grosswendt2007descriptors}. This approach reveals how charged particles produce localized clusters of ionizations that are critical in causing complex DNA damage—lesions that are often irreparable and can lead to cell death or mutation. In our work, we quantify this effect using the complementary cumulative distribution function
\[
F^*(n) = \sum_{\nu > n} p(\nu)\,,
\]
where \(p(\nu)\) is the probability of observing \(\nu\) ionizations in the nanometer target volume. \(F^*(n)\) captures the probability of observing ionization clusters larger than size \(n\).

By focusing on \(F^*(n)\), we provide a sensitive descriptor of ionization events likely to induce complex DNA damage. Larger cluster sizes correlate with increased biological severity, such as double‑strand breaks, which cannot be adequately captured by average descriptors like linear energy transfer (LET). This makes the approach especially relevant across particle types and energies, as large clusters are more frequent in high‑LET radiation such as carbon ions. Overall, nanodosimetry provides a window into the microscopic structure of radiation tracks and their biological impact. Quantifying the frequency of ionization clusters of a given size yields mechanistic insight into radiation-induced DNA lesions, thereby enhancing the interpretability and predictive power of radiobiological models for cell survival. It has been further recently shown that \(F^*(n)\), for \(n = 1, 2, 3\), correlates with survival probability \cite{conte2024nanodosimetry}. Furthermore, experimental studies have demonstrated that cumulative probabilities of observing at least two or three ionizations in nanometer-sized volumes are proportional to inactivation cross-sections at defined survival levels across different cell lines and radiation qualities \cite{selva2019nanodosimetry}. Advanced Monte Carlo track-structure simulations have also validated these distributions and their connection to biological endpoints, supporting the use of cluster‑dose metrics in treatment planning \cite{faddegon2023ionization, ortiz2025nanodosimetry, hilgers2024trackstructure}.

\subsection{Microdosimetry}

Microdosimetry aims to describe the stochastic nature of energy deposition at the micrometer scale, which is relevant to cellular and subcellular structures such as nuclei and chromatin domains. Unlike nanodosimetry, which focuses on ionization clusters within DNA-sized volumes, microdosimetry characterizes the distribution of energy imparted to volumes comparable to whole cells or organelles, providing insight into the probability of inducing clustered DNA damage and its repair dynamics. This approach is essential for understanding biological endpoints such as clonogenic survival, as it bridges the gap between track-structure physics and radiobiological response at the cellular level \cite{icru2016report, rossi1996microdosimetry}.
In this study, microdosimetric quantities were simulated using the TOPAS Monte Carlo toolkit (version 3.9) \cite{perl2012topas}, which has been validated against experimental measurements for a variety of radiation qualities \cite{zhu2019microdosimetric, missiaggia2023investigation, burigo2014microdosimetry}. Radiation fields were generated to replicate the experimental conditions reported in PIDE, with particles directed from an environmental source toward the center of the simulated geometry. Energy deposition was scored in a water sphere of 1$\mu$m diameter, representing a typical cell nucleus.
The primary descriptors considered are the frequency-mean lineal energy $\bar{y}_F$ and dose-mean lineal energy $\bar{y}_D$, defined as:
\[
\bar{y}_F = \int_0^\infty y f(y) dy\,,\quad \bar{y}_D = \frac{1}{\bar{y}_F} \int_0^\infty y^2 f(y) dy\,,
\]
where $f(y)$ is the probability density function of lineal energy $y$, expressed in keV/$\mu$m. The quantity $\bar{y}_F$ represents the average energy deposition per unit length for individual events, while $\bar{y}_D$ weights higher-energy events more strongly, making it particularly relevant for assessing biological effectiveness. These metrics have been historically linked to RBE and cell survival, as they capture the heterogeneity of energy deposition at the micrometer scale, which influences the induction of complex DNA damage and subsequent repair processes \cite{icru2016report, rossi1996microdosimetry}.

By incorporating microdosimetric descriptors alongside nanodosimetric quantities, we enable a multiscale characterization of radiation quality, providing a more comprehensive basis for predicting biological outcomes such as cell survival. This integration is crucial because different spatial scales, nanometer for DNA damage induction and micrometer for cellular energy deposition, jointly determine the probability of cell inactivation. Microdosimetry is currently implemented in clinical carbon ion therapy, particularly in Japan, where it forms the basis of the MKM used to calculate relative biological effectiveness (RBE) for treatment planning \cite{hawkins1994cell, inaniwa2010treatment, bellinzona2021linking}.

\subsection{Deep Learning with attention mechanism}
The developed model is based on \textit{TabNet}~\cite{arik2021tabnet}, a deep learning architecture specifically designed for tabular data. It employs a sequential attention mechanism that enables adaptive feature selection at each decision step, achieving high predictive performance while providing intrinsic interpretability.

The model operates through a sequence of $T$ decision steps. At each step, an attentive transformer produces a sparse feature selection mask that determines which input features are used at that step. The masked features are then processed by a feature transformer, composed of fully connected layers with Gated Linear Units (GLUs), to generate higher-level representations. The feature transformer is composed of shared and decision-step–specific blocks, enabling parameter sharing across steps while allowing step-wise specialization. At each decision step, an attentive transformer computes a sparse feature mask using Sparsemax, conditioned on both the transformed features from the previous step and a feature prior that discourages repeated feature usage across steps. This prior is updated iteratively, promoting diverse feature selection and enhancing interpretability. The masked input is then processed by the feature transformer, producing both decision outputs and representations for subsequent steps. A sparsity regularization term based on the entropy of the attention masks is added to the training objective.

The architecture consists of four main components: (i) Feature Transformer, (ii) Attentive Transformer, (iii) Feature Prior and Decision Steps, and (iv) Sparse Feature Selection and Regularization.

\paragraph{Feature Transformer.}
Let
\[
\mathbf{X} \in \mathbb{R}^{n \times d}
\]
denote the input matrix, where $n$ is the number of samples and $d$ is the number of features. The feature transformer maps masked input features into higher-level representations using a stack of fully connected layers followed by GLU. It consists of a set of shared layers, common to all decision steps, and step-specific layers that allow specialization at each step.

A generic transformation can be written as
\[
\mathbf{H} = \mathrm{GLU}\!\left( \mathbf{X} \mathbf{W}_1 + \mathbf{b}_1 \right),
\qquad
\mathbf{H}' = \mathrm{GLU}\!\left( \mathbf{H} \mathbf{W}_2 + \mathbf{b}_2 \right),
\]
where $\mathbf{W}_1, \mathbf{W}_2$ are learnable weight matrices and $\mathbf{b}_1, \mathbf{b}_2$ are bias vectors.

The Gated Linear Unit is defined as
\[
\mathrm{GLU}(\mathbf{Z}) = \mathbf{Z}_a \odot \sigma(\mathbf{Z}_b),
\]
where $\mathbf{Z} \in \mathbb{R}^{n \times 2h}$ is split into two equal parts $\mathbf{Z}_a, \mathbf{Z}_b \in \mathbb{R}^{n \times h}$, $\odot$ denotes element-wise multiplication, and $\sigma(\cdot)$ is the sigmoid activation function. This gating mechanism enables adaptive control of information flow and improves model stability.

\paragraph{Attentive Transformer and Feature Prior.}
At each decision step $t$, the attentive transformer computes a sparse feature selection mask $\mathbf{M}^{(t)} \in \mathbb{R}^{n \times d}$ using the Sparsemax activation function. The mask is conditioned on the transformed features from the previous step and a feature prior $\mathbf{P}^{(t)}$, which controls feature reuse across decision steps:
\[
\mathbf{M}^{(t)} =
\mathrm{Sparsemax}\!\left( \mathbf{A}^{(t)} \odot \mathbf{P}^{(t)} \right),
\]
where $\mathbf{A}^{(t)}$ denotes the output of the attentive transformer and $\odot$ represents element-wise multiplication.

The feature prior is initialized as $\mathbf{P}^{(1)} = \mathbf{1}$ and updated after each decision step according to
\[
\mathbf{P}^{(t+1)} = \gamma \left( \mathbf{P}^{(t)} - \mathbf{M}^{(t)} \right),
\qquad \gamma > 1,
\]
thereby discouraging repeated selection of the same features and promoting diversity across decision steps.

The Sparsemax function projects its input onto the probability simplex:
\[
\mathrm{Sparsemax}(\mathbf{z}) =
\arg\min_{\mathbf{p} \in \Delta^d}
\|\mathbf{p} - \mathbf{z}\|_2^2,
\]
where $\Delta^d = \{ \mathbf{p} \in \mathbb{R}^d \mid \mathbf{p} \ge 0,\ \sum_{j=1}^d p_j = 1 \}$. Unlike Softmax, Sparsemax produces sparse probability distributions, enabling explicit feature selection.

\paragraph{Decision Steps and Aggregation.}
At decision step $t$, the input features are masked as
\[
\mathbf{X}^{(t)} = \mathbf{M}^{(t)} \odot \mathbf{X},
\]
and passed through the feature transformer to produce a latent representation
\[
\mathbf{H}^{(t)} = \mathrm{FeatureTransformer}\!\left( \mathbf{X}^{(t)} \right).
\]
Each step yields a decision output $P^{(t)}$, and the final model prediction is obtained by aggregating all decision step outputs:
\[
P = \sum_{t=1}^{T} P^{(t)}.
\]

\paragraph{Sparse Regularization.}
To further encourage sparse and interpretable feature selection, a regularization term based on the entropy of the attention masks is included
\[
\mathcal{L}_{\mathrm{reg}} =
\frac{\lambda}{T} \sum_{t=1}^{T}
\sum_{j=1}^{d}
M^{(t)}_{j}
\log\!\left( M^{(t)}_{j} + \epsilon \right),
\]
where $\lambda$ controls the regularization strength and $\epsilon$ is a small constant for numerical stability. This term penalizes dense feature usage, enhancing interpretability by encouraging the model to focus on a limited subset of features at each decision step.

\begin{figure}[htbp]
    \centering
    \includegraphics[page=1,width=\linewidth]{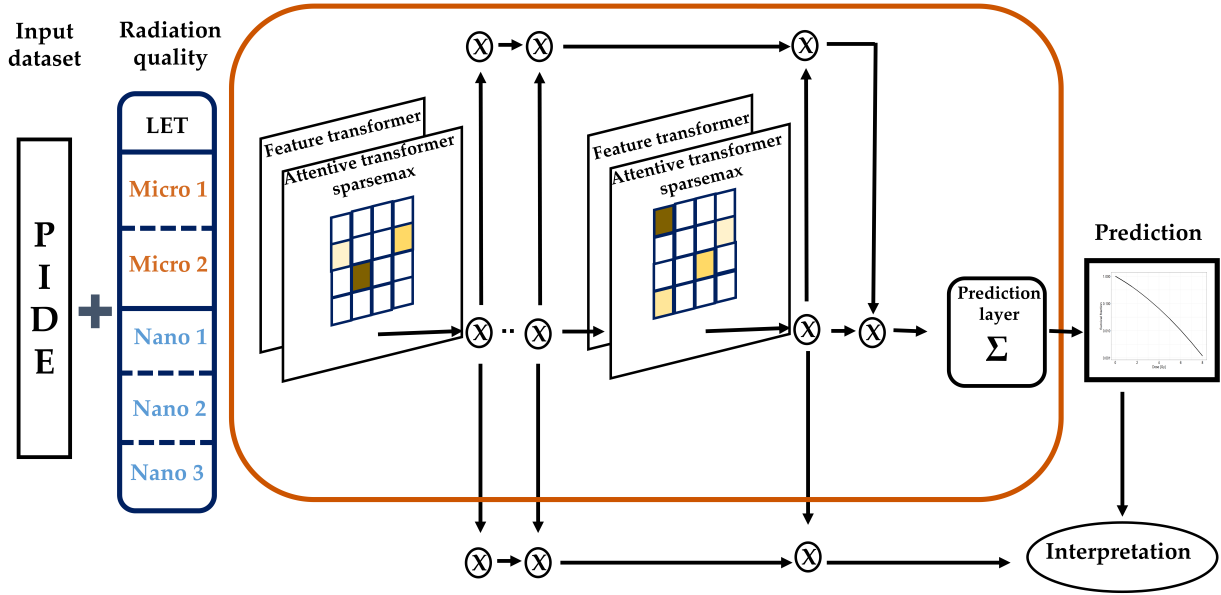}
    \caption{DL model workflow.}
    \label{fig:Scheme}
\end{figure}

Figure~\ref{fig:Scheme} illustrates the overall architecture of the model, from dataset construction using PIDE to the sequential attention-based decision steps leading to the final interpretable predictions.
\subsection{Error analysis and biological endpoints considered}

All the selected data from PIDE, where only the LET is reported, are augmented with the nano- and microdosimetric quantities together with dose and surviving fraction values. This dataset also contains the biological parameters of each experiment in terms of the linear-quadratic parameters for photons $\alpha_X$ and $\beta_X$. The final structure of the complete dataset is shown in Table 1.
Performance is assessed using Mean Absolute Percentage Error (MAPE) and Root Mean Square Error (RMSE). The latter is defined for each SF experiment as

\[
\log \mathrm{RMSE}_i := \sqrt{\frac{1}{N_D} \sum_{D} \bigl( \log(\hat{S}_i(D)) - \log(S_i(D)) \bigr)^2 }\,,
\]

where $N_D$ is the number of doses measured in the $i$-th experiment, while $\hat{S}_i$ and $S_i$ are the cell survivals predicted and measured, respectively. On the other hand, given N observations, true values $y_i$, and predictions $\hat{y}_i$, MAPE can be defined as
\[
\mathrm{MAPE} := \frac{1}{N} \sum_{i=0}^{N} \left| \frac{\mathrm{RBE}_i - \hat{\mathrm{RBE}}_i}{\mathrm{RBE}_i} \right| \times 100\,.
\]
This value is computed from the predictions of $\mathrm{RBE}_{10}$, defined in equation~\eqref{EQN:RBE}, where the biological endpoint corresponds to a survival probability of $10\%$.

Three different biological endpoints are considered throughout the paper to study the attention masks: (i) the dose corresponding to 10$\%$ survival probability, (ii) dose 2 Gy, and (iii) the low dose limits. Regarding this last endpoint, we consider the average of the attention at doses 0.25, 0.5, and 0.75 Gy. All importance values reported throughout the manuscript are normalized by the number of experiments in a given radiation field condition, ensuring that conditions with more replicates do not disproportionately influence the final importance scores.

Importance curves against the energy of the beam for protons and carbon ions are generated using Locally Estimated Scatterplot Smoothing (LOESS) \cite{cleveland1979robust}, a non-parametric method for smoothing a series of data, and also constructing confidence intervals around the curve. 

%\subsection*{Data Preprocessing}
%
%\begin{table}[h!]
%\centering
%\begin{tabular}{l c}
%\hline
%\textbf{Parameter} & \textbf{Value} \\
%\hline
%Epochs & 100 \\
%Batch size & 64 \\
%Validation split & 20\% \\
%Learning rate & 0.002 \\
%Number of attention steps & 3 \\
%Sparsity regularizer & 0.001 \\
%\hline
%\end{tabular}
%\caption{Main parameters of the model.}
%\end{table}

Analyses were conducted with Rversion4.4.2 with torch(0.16.0), tidyverse(2.0.0), rUM(2.2.0), and gtsummary(2.4.0) packages \cite{R-base, R-tidyverse, tidyverse2019, ggplot22016, R-rUM}.

\section{Results}

\begin{figure}[htbp]
  \centering
  \begin{subfigure}[t]{0.48\textwidth}
    \centering
    \includegraphics[page=1,width=\linewidth]{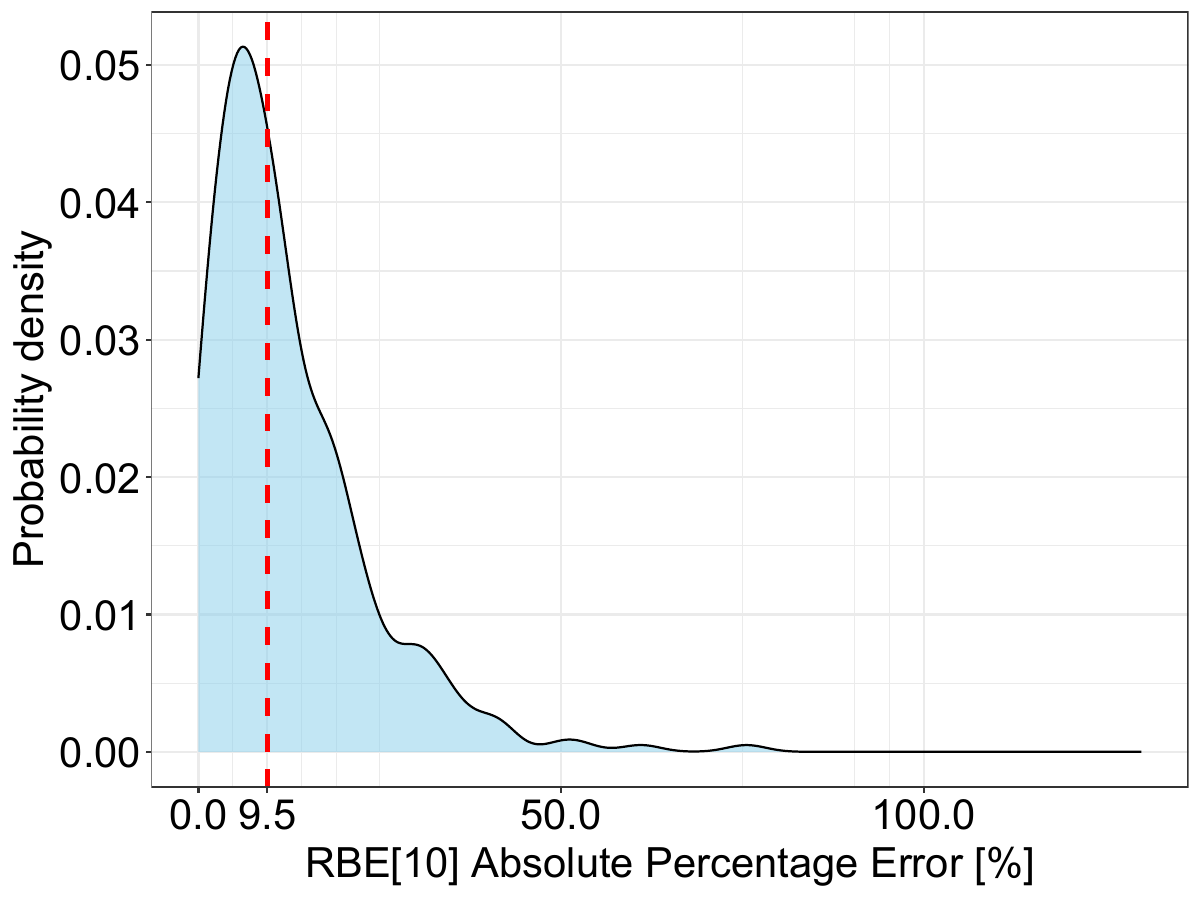}
    \caption{Absolute percentage error from model’s predictions. The red dotted line indicates the MAPE.}
    \label{fig:MAPE}
  \end{subfigure}
  \hfill
  \begin{subfigure}[t]{0.48\textwidth}
    \centering
    \includegraphics[page=1,width=\linewidth]{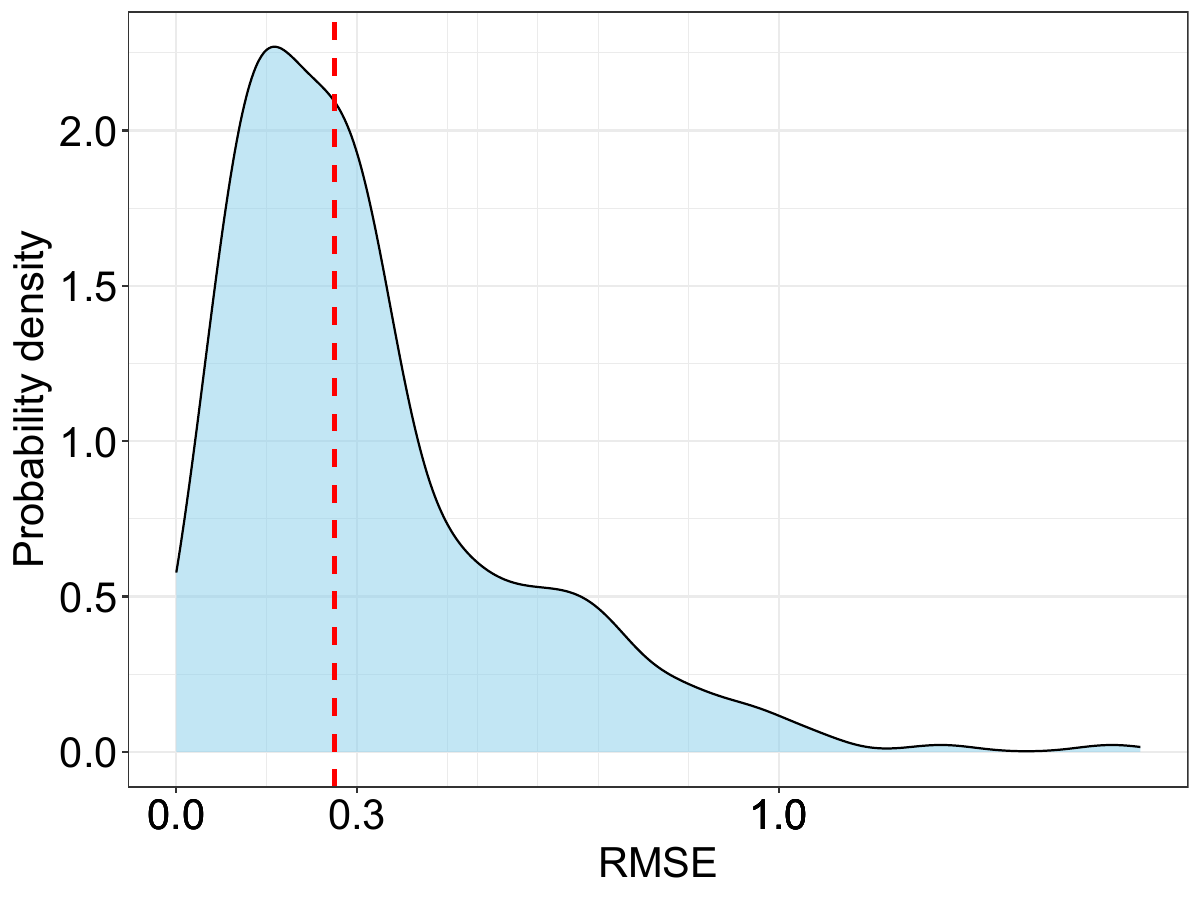}
    \caption{Root mean square error. The red dotted line indicates the average value of the distribution.}
    \label{fig:RMSE}
  \end{subfigure}
  \hfill
  \begin{subfigure}[t]{0.48\textwidth}
    \centering
    \includegraphics[page=1,width=\linewidth]{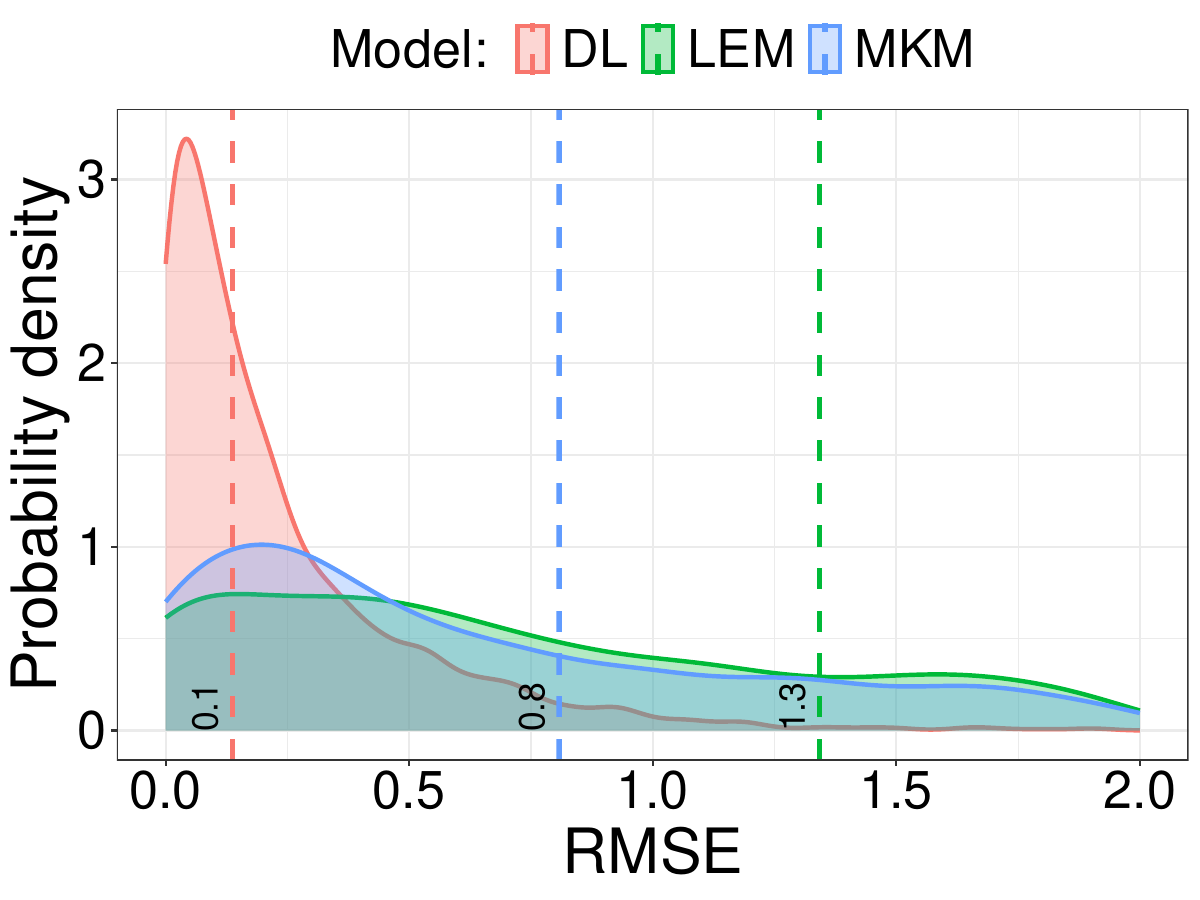}
    \caption{Root mean square error for our model compared with LEM and MKM on data from V79 cell line. The dotted lines indicate the average value of the distribution for each model.}
    \label{fig:RMSE_V79}
  \end{subfigure}
  \hfill
  \begin{subfigure}[t]{0.48\textwidth}
    \centering
    \includegraphics[page=1,width=\linewidth]{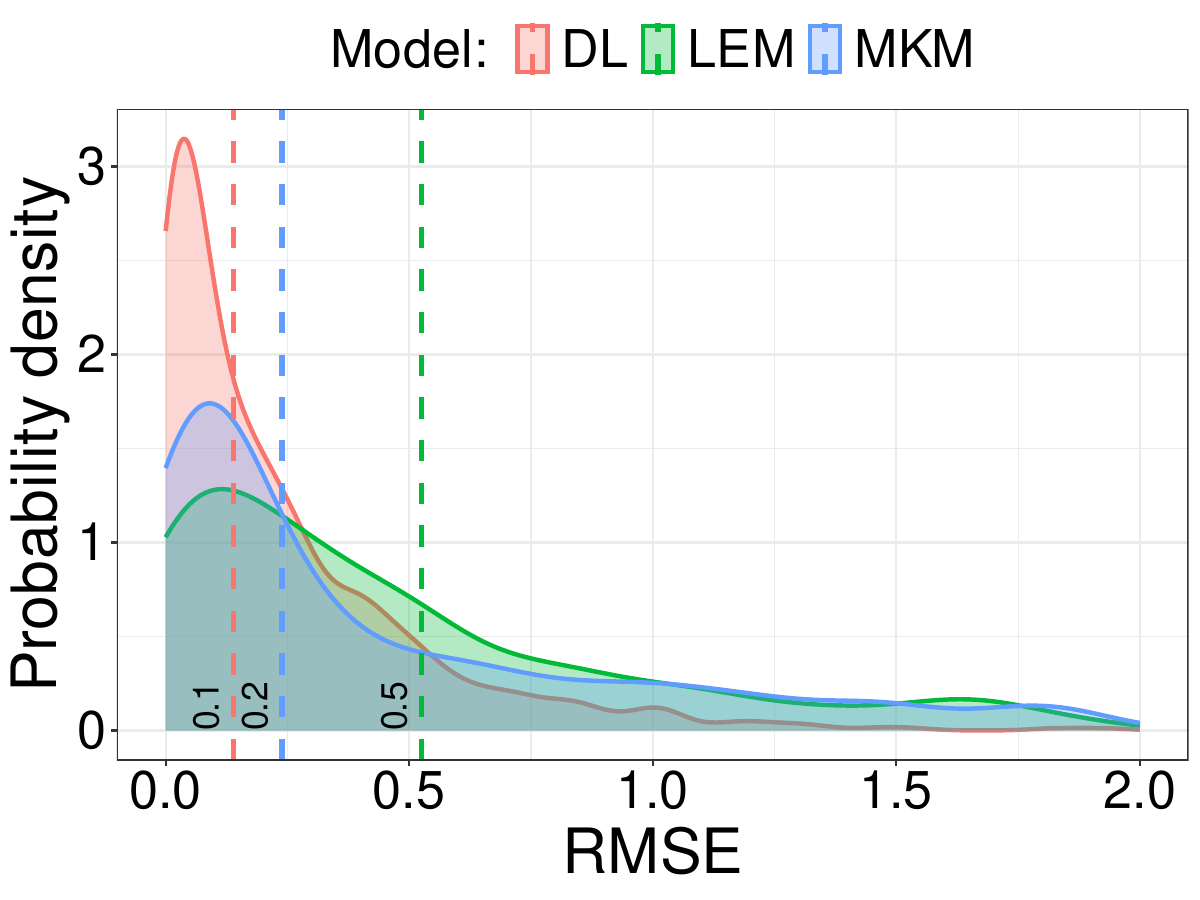}
    \caption{Root mean square error for our model compared with LEM and MKM on data from HSG cell line. The dotted lines indicate the average value of the distribution for each model.}
    \label{fig:RMSE_HSG}
  \end{subfigure}
  \caption{Model performance.}
  \label{fig:Err}
\end{figure}

The accuracy of the RBE$_{10}$ model predictions can be evaluated by absolute percentage error distribution, Figure \ref{fig:MAPE}, with a MAPE of 9.5$\%$. Figure \ref{fig:RMSE} shows the evaluation also in terms of RMSE distribution, with a mean value of 0.3, Figure \ref{fig:RMSE}. In Figure \ref{fig:RMSE_V79} and \ref{fig:RMSE_HSG} we compare the results of our model with the MKM \cite{kase2007biophysical} and LEM \cite{russo2011analysis} on survival data coming from V79 and HSG cell lines, which are the most represented in the whole dataframe. The multiscale DL model shows in both cases the lowest RSME around 0.1. The performance from MKM is assessed at 0.8 and 0.2 for V79 and HSG respectively, while these values for LEM are 1.3 and 0.5.

\begin{figure}[htbp]
  \centering
  \begin{subfigure}[t]{0.48\textwidth}
    \centering
    \includegraphics[page=1,width=\linewidth]{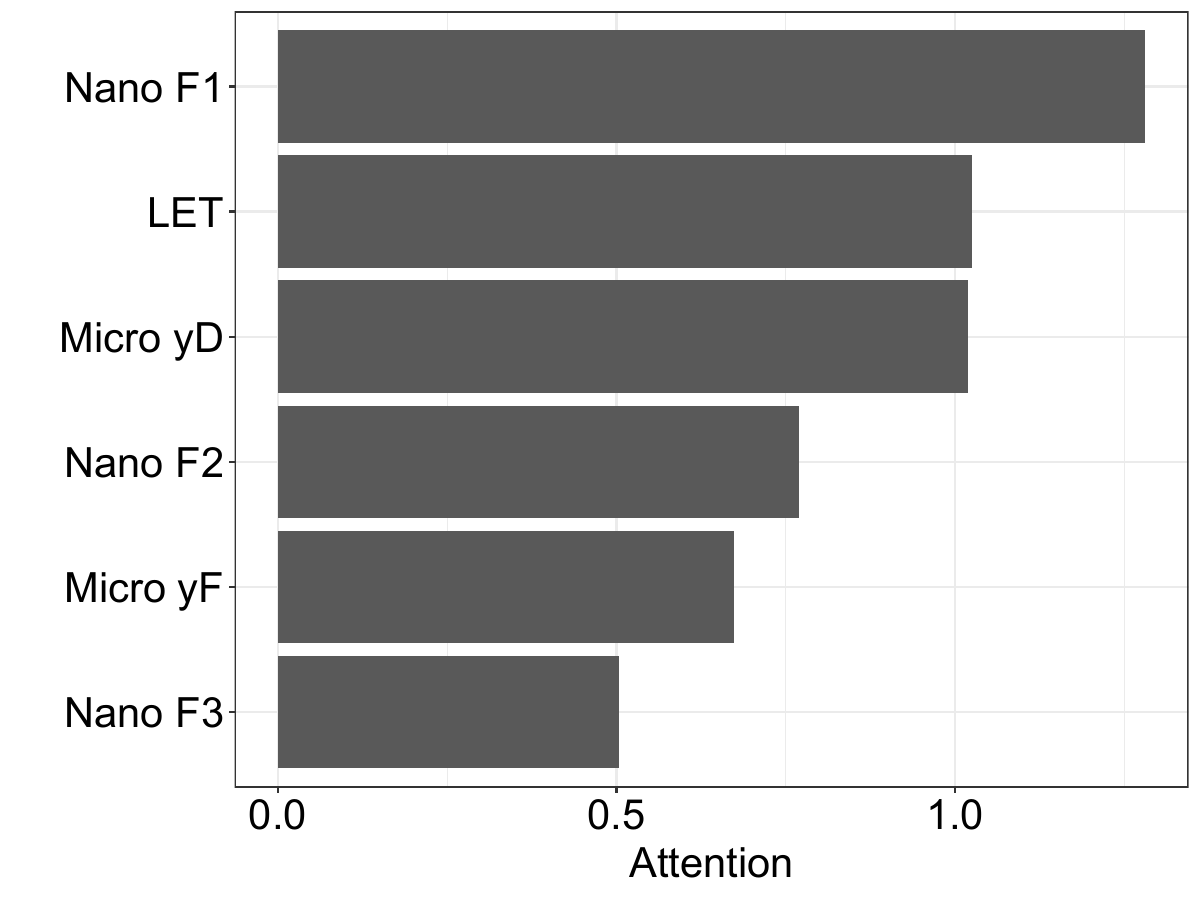}
    \caption{Global attention evaluated at dose corresponding to 10$\%$ survival, extracted from all the attention steps across the entire dataset.}
    \label{fig:Att10}
  \end{subfigure}
  \hfill
  \begin{subfigure}[t]{0.48\textwidth}
    \centering
    \includegraphics[page=1,width=\linewidth]{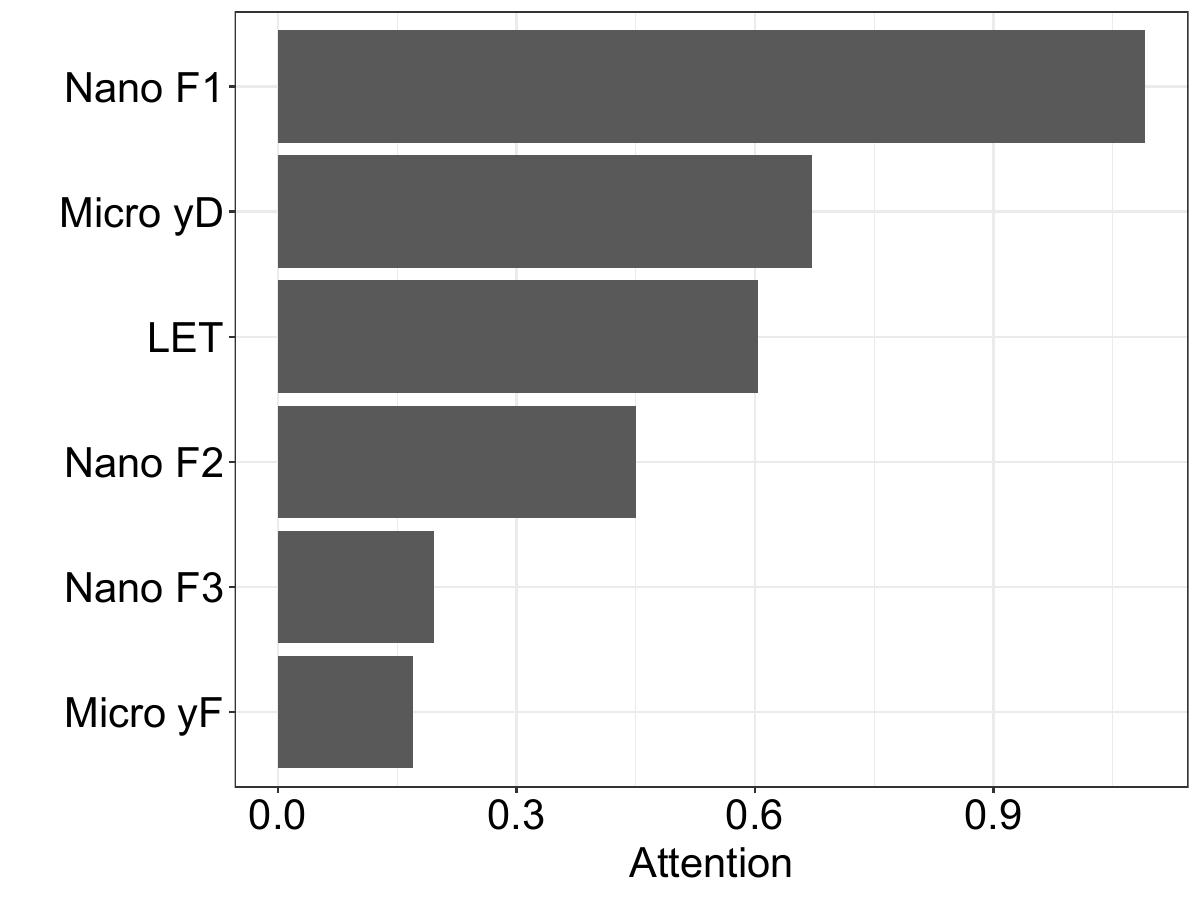}
    \caption{Global attention evaluated at dose 2 Gy, extracted from all the attention steps across the entire dataset.}
    \label{fig:Att2}
  \end{subfigure}
  \hfill
  \begin{subfigure}[t]{0.48\textwidth}
    \centering
    \includegraphics[page=1,width=\linewidth]{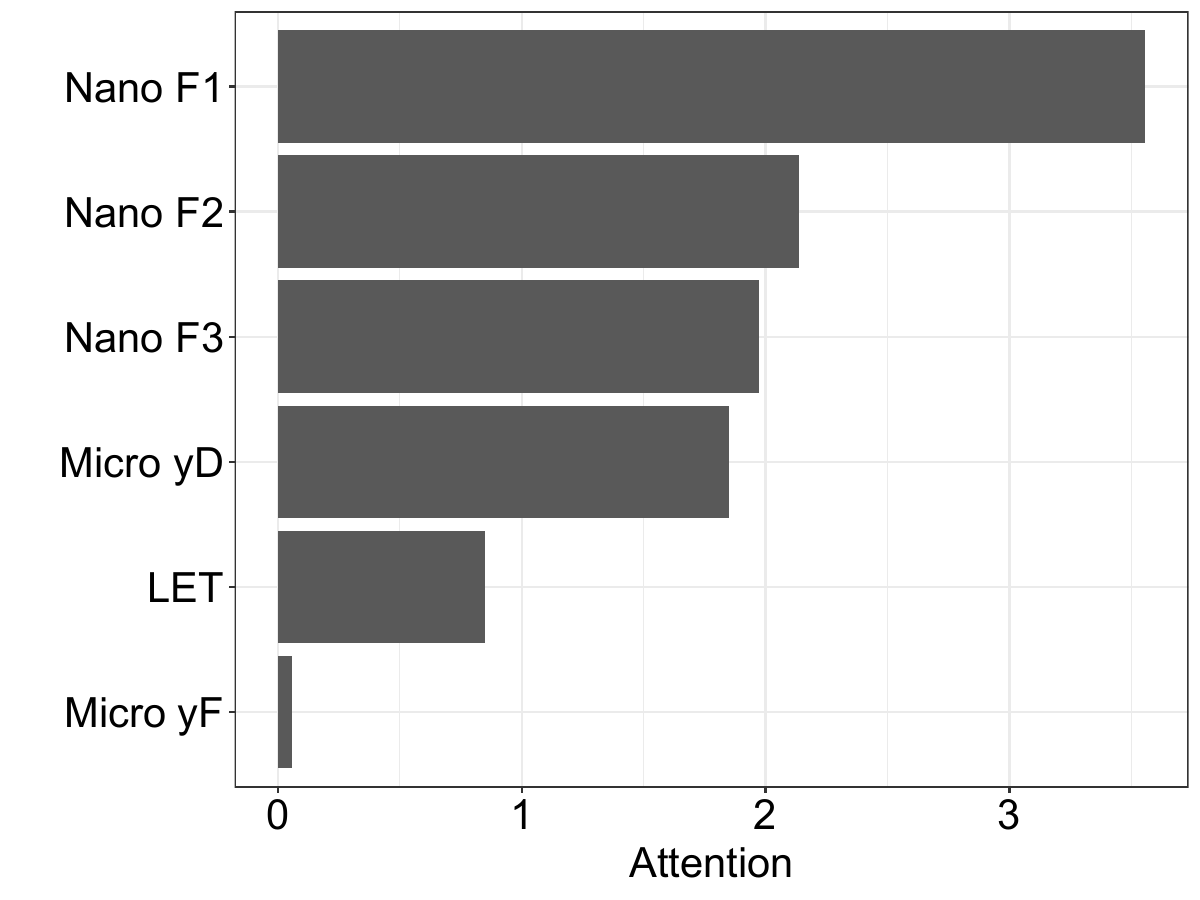}
    \caption{Global attention evaluated at dose corresponding to the low dose limit, extracted from all the attention steps across the entire dataset.}
    \label{fig:AttLD}
  \end{subfigure}
  \caption{Global attention for the three endpoints considered.}
  \label{fig:Att}
\end{figure}

We assessed the contribution of each feature using the built-in interpretability of the attention mechanism at each prediction step. Global feature importance is summarized in Figure~\ref{fig:Att}, with detailed results for $D_{10}$ in Figure~\ref{fig:Att10}, for survival at 2~Gy in Figure~\ref{fig:Att2}, and for the low-dose limit in Figure~\ref{fig:AttLD}. Across all endpoints, the nanodosimetric descriptor $F^*(1)$ consistently exhibits the highest attention, indicating its dominant role in survival prediction. For $D_{10}$ and 2~Gy, attention is more evenly distributed, with $\bar{y}_D$ and LET emerging as the most influential features after $F^*(1)$. In contrast, the low-dose limit shows a clear predominance of nanodosimetric metrics, with only $\bar{y}_F$ approaching comparable importance. These results highlight the varying relevance of physical descriptors across biological endpoints, with nanodosimetry playing a particularly critical role in the low-dose region.

\begin{figure}[htbp]
  \centering
  \begin{subfigure}[t]{0.48\textwidth}
    \centering
    \includegraphics[page=1,width=\linewidth]{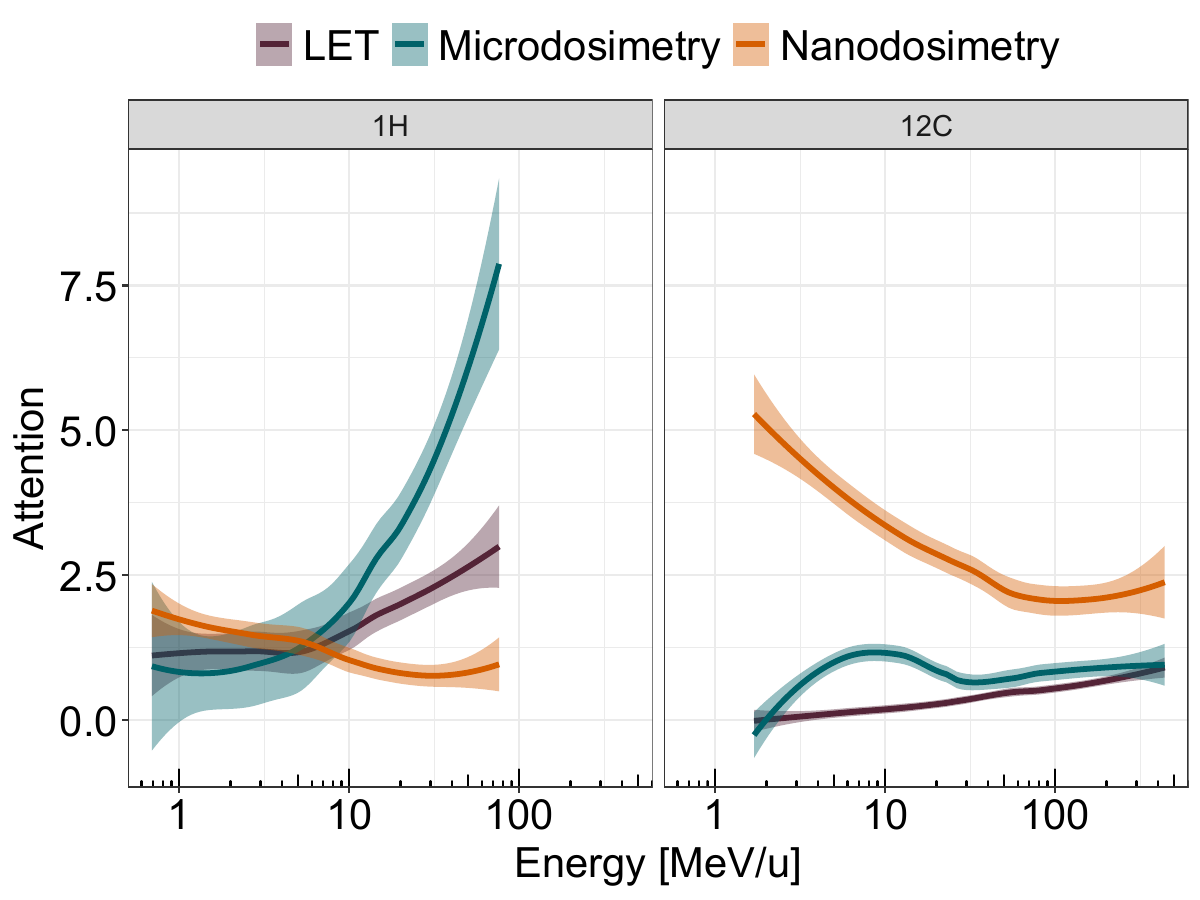}
    \caption{Attention to LET, microdosimetric, and nanodosimetric features evaluated at dose corresponding to 10$\%$ survival as a function of beam energy for protons and carbon ions. Shaded areas indicate 95$\%$ confidence intervals with the LOESS method.}
    \label{fig:EngAtt10}
  \end{subfigure}
  \hfill
  \begin{subfigure}[t]{0.48\textwidth}
    \centering
    \includegraphics[page=1,width=\linewidth]{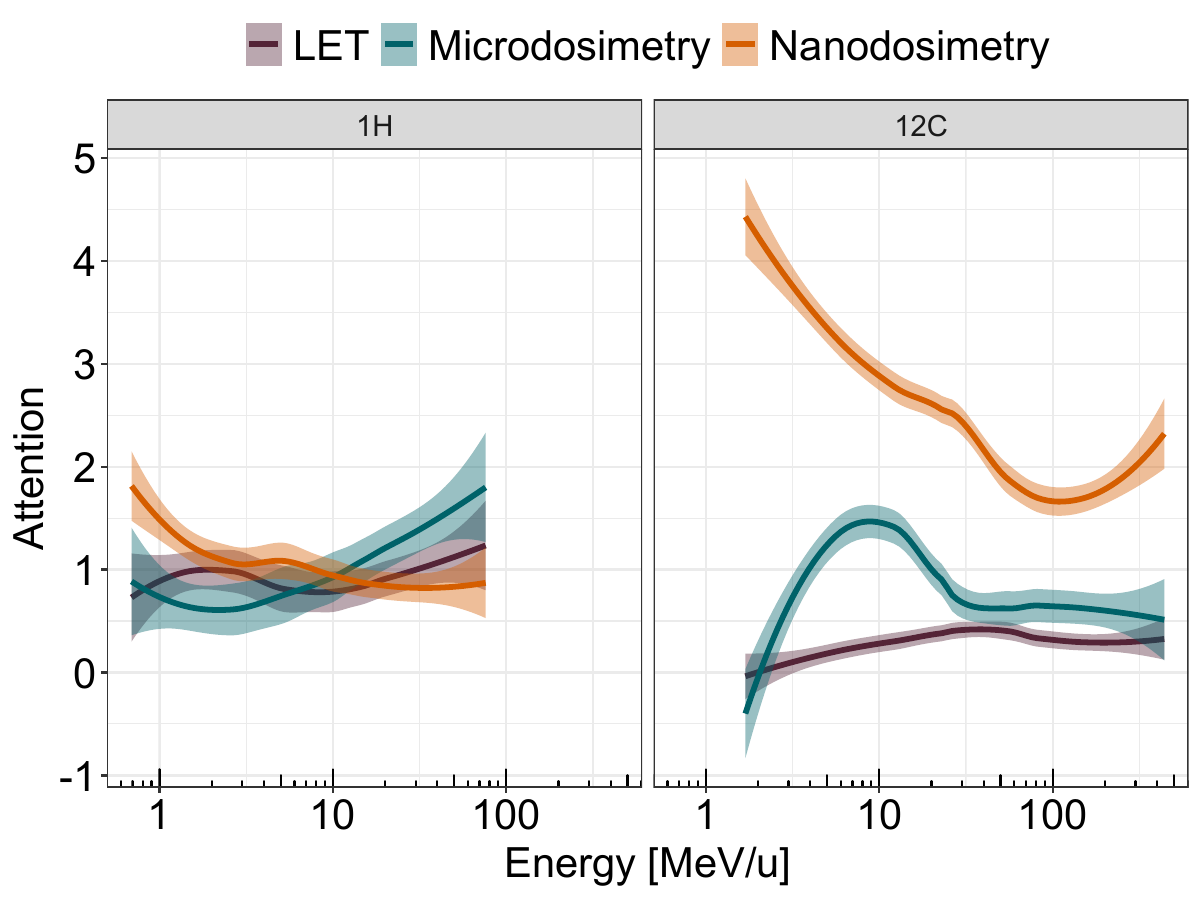}
    \caption{Attention to LET, microdosimetric, and nanodosimetric features evaluated at 2 Gy as a function of beam energy for protons and carbon ions. Shaded areas indicate 95$\%$ confidence intervals with the LOESS method.}
    \label{fig:EngAtt2}
  \end{subfigure}
  \hfill
  \begin{subfigure}[t]{0.48\textwidth}
    \centering
    \includegraphics[page=1,width=\linewidth]{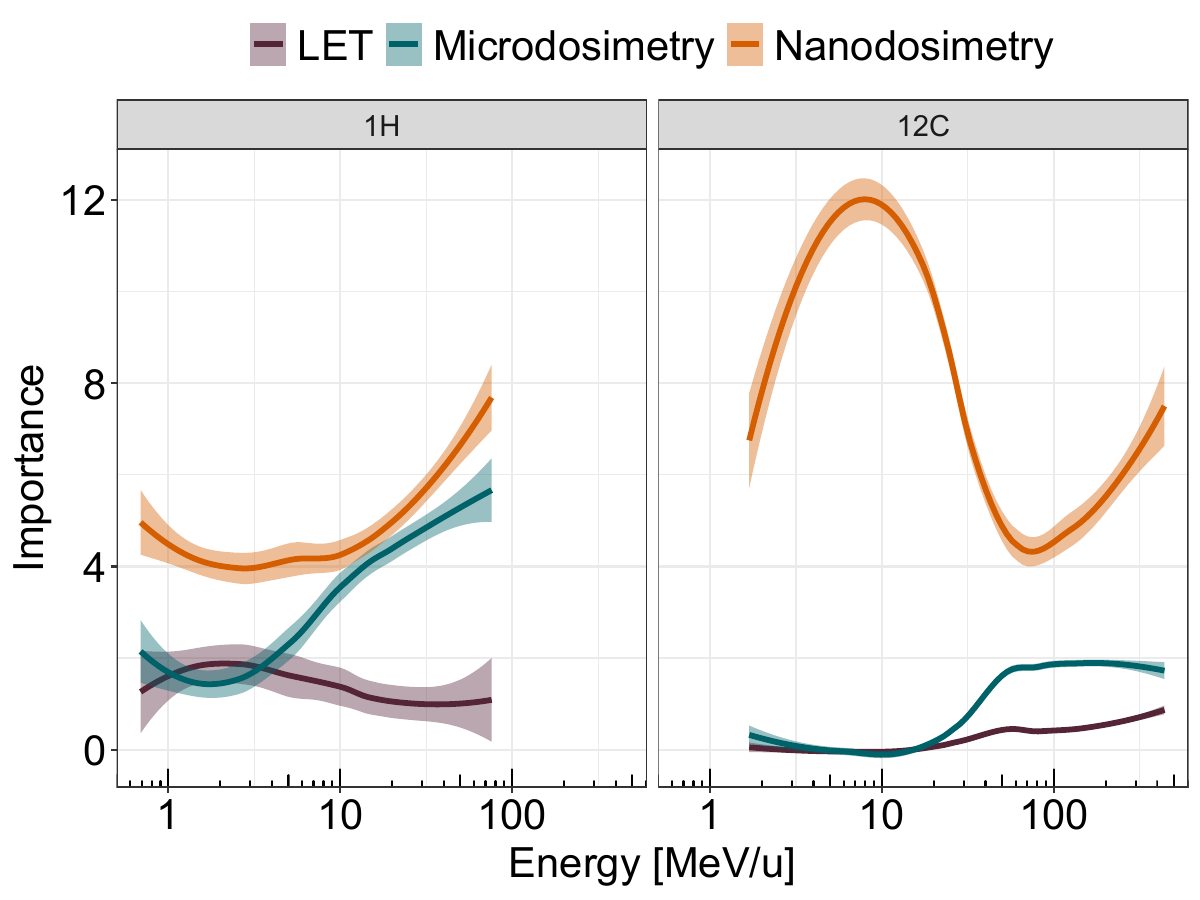}
    \caption{Attention to LET, microdosimetric, and nanodosimetric features evaluated at dose corresponding to the low dose limit as a function of beam energy for protons and carbon ions. Shaded areas indicate 95$\%$ confidence intervals with the LOESS method.}
    \label{fig:EngAttLD}
  \end{subfigure}
  \caption{Attention as a function of beam energy for protons and carbon ions for the three endpoints considered.}
  \label{fig:EngAtt}
\end{figure}

While global importance offers a broad overview, our goal was to achieve a deeper understanding by analyzing local importance, which captures condition-specific contributions to model predictions. Figure~\ref{fig:EngAtt} illustrates attention as a function of beam energy, separated for protons and carbon ions, with detailed results for $D_{10}$ in Figure~\ref{fig:EngAtt10}, survival at 2~Gy in Figure~\ref{fig:EngAtt2}, and the low-dose limit in Figure~\ref{fig:EngAttLD}. Different spatial scales are represented by distinct colors.
For protons, in the case of $D_{10}$ and survival at 2~Gy, microdosimetry and LET exhibit similar trends, generally increasing with energy, whereas nanodosimetry shows the opposite behavior. Microdosimetry becomes significantly more important at higher proton energies for $D_{10}$, while attention is more uniformly distributed for survival at 2~Gy. In contrast, LET has minimal importance for carbon ion irradiation. However, its trend remains broadly similar to microdosimetry, except for a small peak around 10~MeV/u. Nanodosimetry displays a globally decreasing tendency with increasing beam energy for carbon ions, but its relative importance compared to higher-level scales is striking—approximately three times greater than for protons.
For the low-dose limit, nanodosimetry emerges as the dominant contributor for both protons and carbon ions, with its attention in carbon ions surpassing all other scales. LET maintains a nearly constant importance for both particle types, whereas microdosimetry shows an increasing trend with energy. Nanodosimetry, on the other hand, reveals an upward trend at higher proton energies and a complex pattern for carbon ions, characterized by a pronounced peak near 10~MeV/u.
The shaded regions around the smoothed importance curves represent 95$\%$ confidence intervals, reflecting uncertainty in the mean predicted importance across different radiosensitivities of the cell lines included in the dataset.

\begin{figure}[htbp]
  \centering
  \begin{subfigure}[t]{0.48\textwidth}
    \centering
    \includegraphics[page=1,width=\linewidth]{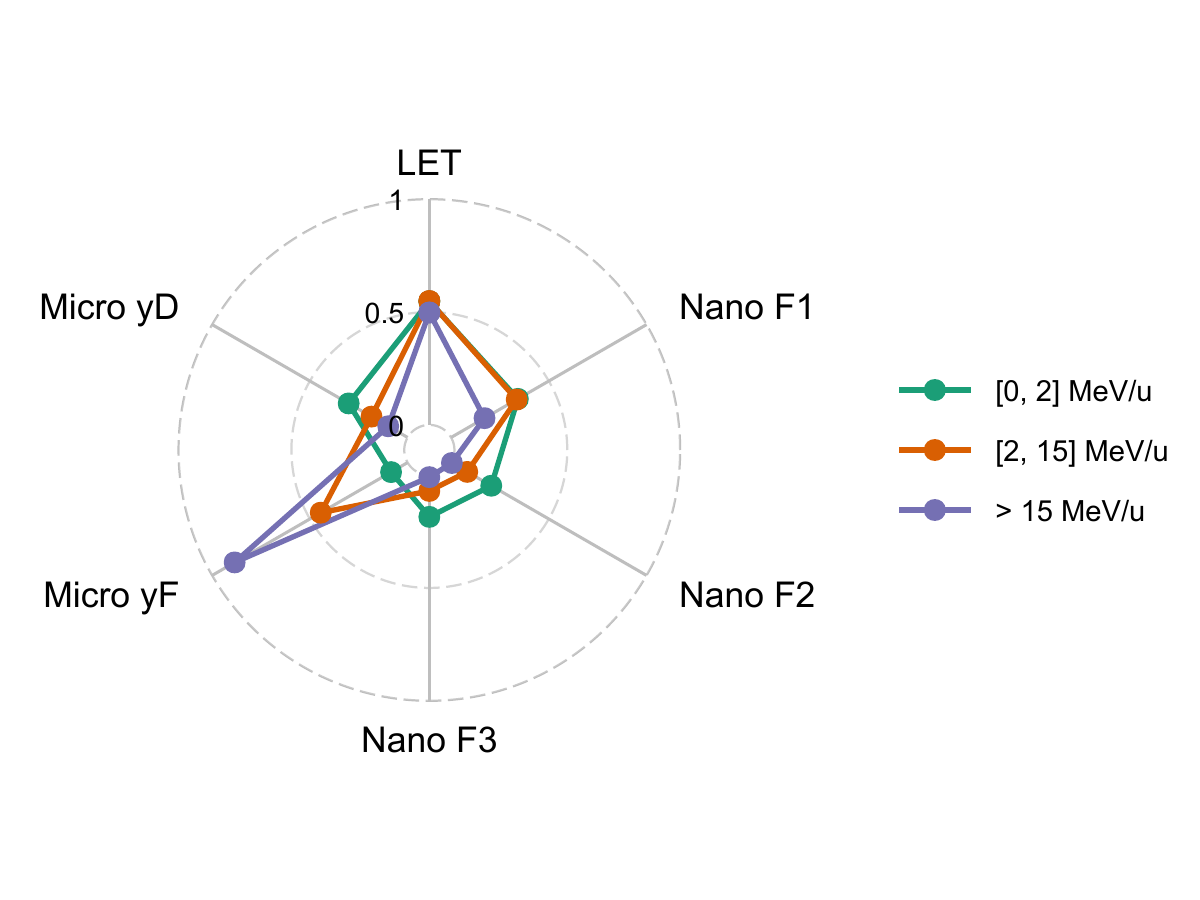}
    \caption{Spider plot with attention for protons at dose corresponding to 10$\%$ survival. Different colors represent different energy bins.}
    \label{fig:Rad10p}
  \end{subfigure}
  \hfill
  \begin{subfigure}[t]{0.48\textwidth}
    \centering
    \includegraphics[page=1,width=\linewidth]{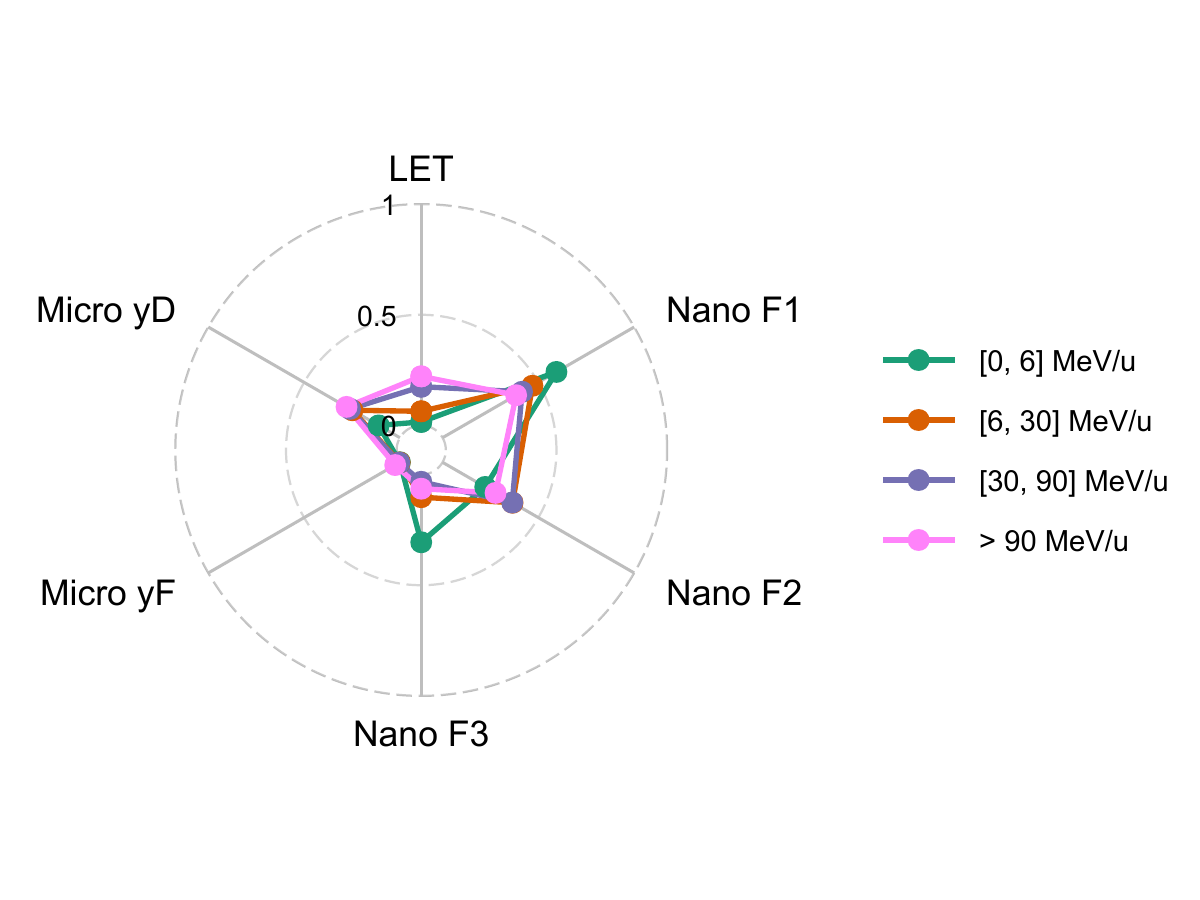}
    \caption{Spider plot with attention for carbon ions at dose corresponding to 10$\%$ survival. Different colors represent different energy bins.}
    \label{fig:Rad10c}
  \end{subfigure}
  \hfill
  \begin{subfigure}[t]{0.48\textwidth}
    \centering
    \includegraphics[page=1,width=\linewidth]{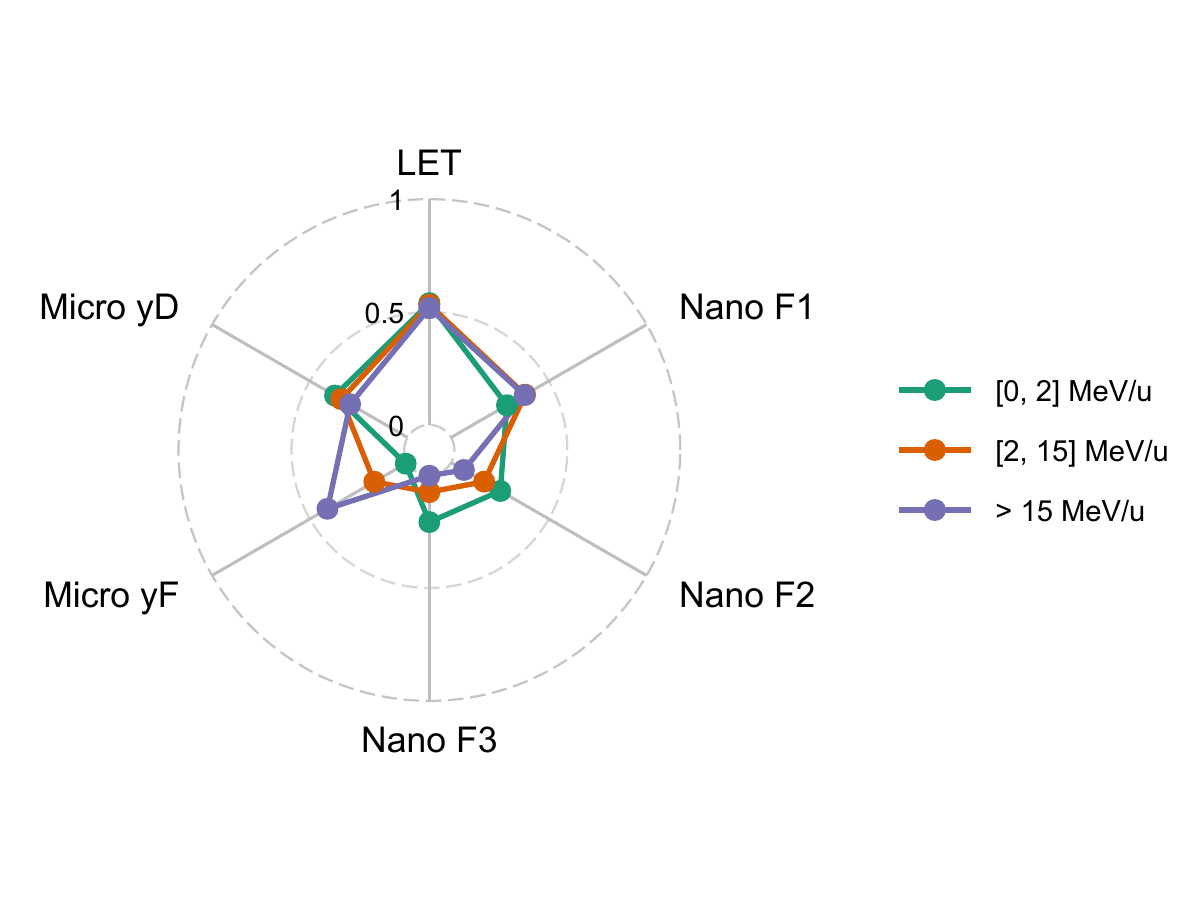}
    \caption{Spider plot with attention for protons at dose 2 Gy. Different colors represent different energy bins.}
    \label{fig:Rad2p}
  \end{subfigure}
  \hfill
  \begin{subfigure}[t]{0.48\textwidth}
    \centering
    \includegraphics[page=1,width=\linewidth]{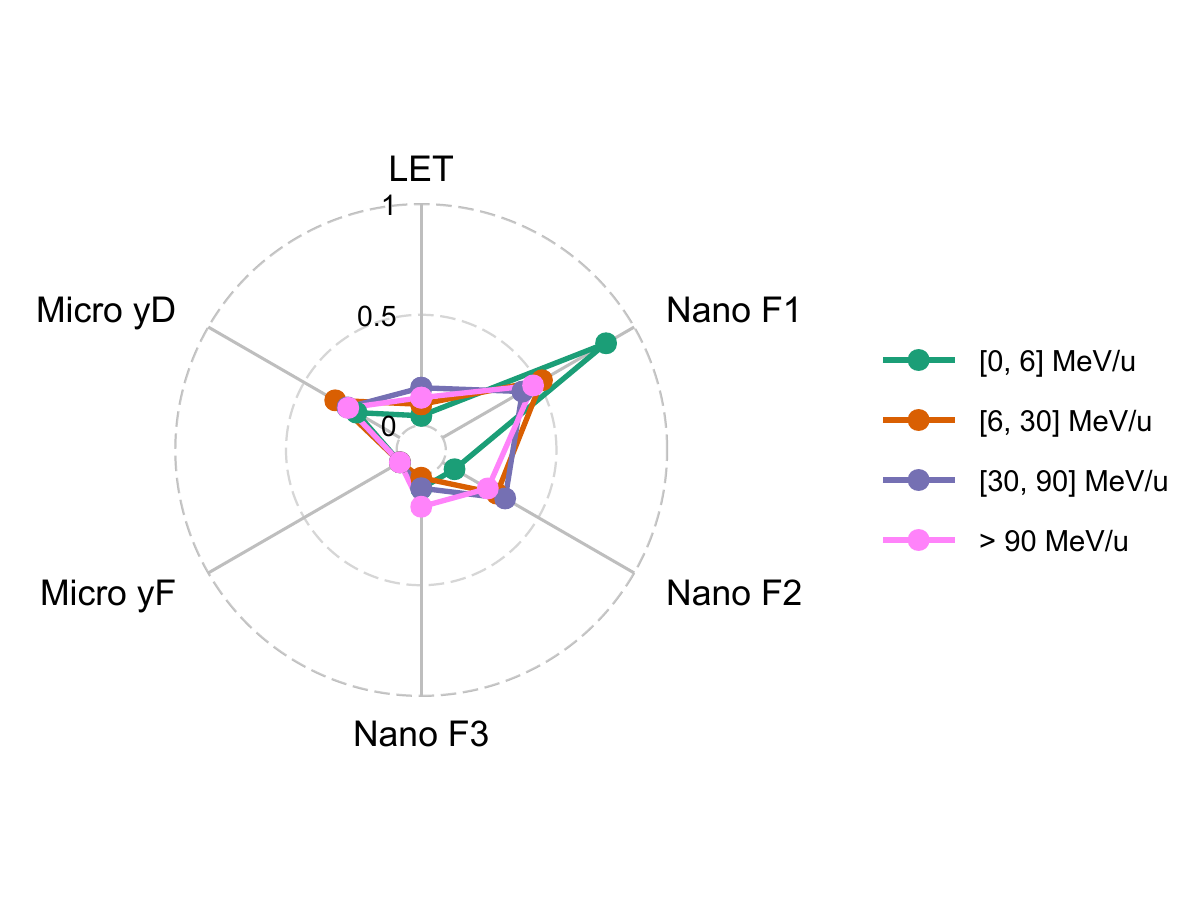}
    \caption{Spider plot with attention for carbon ions at dose 2 Gy. Different colors represent different energy bins.}
    \label{fig:Rad2C}
  \end{subfigure}
    \hfill
  \begin{subfigure}[t]{0.48\textwidth}
    \centering
    \includegraphics[page=1,width=\linewidth]{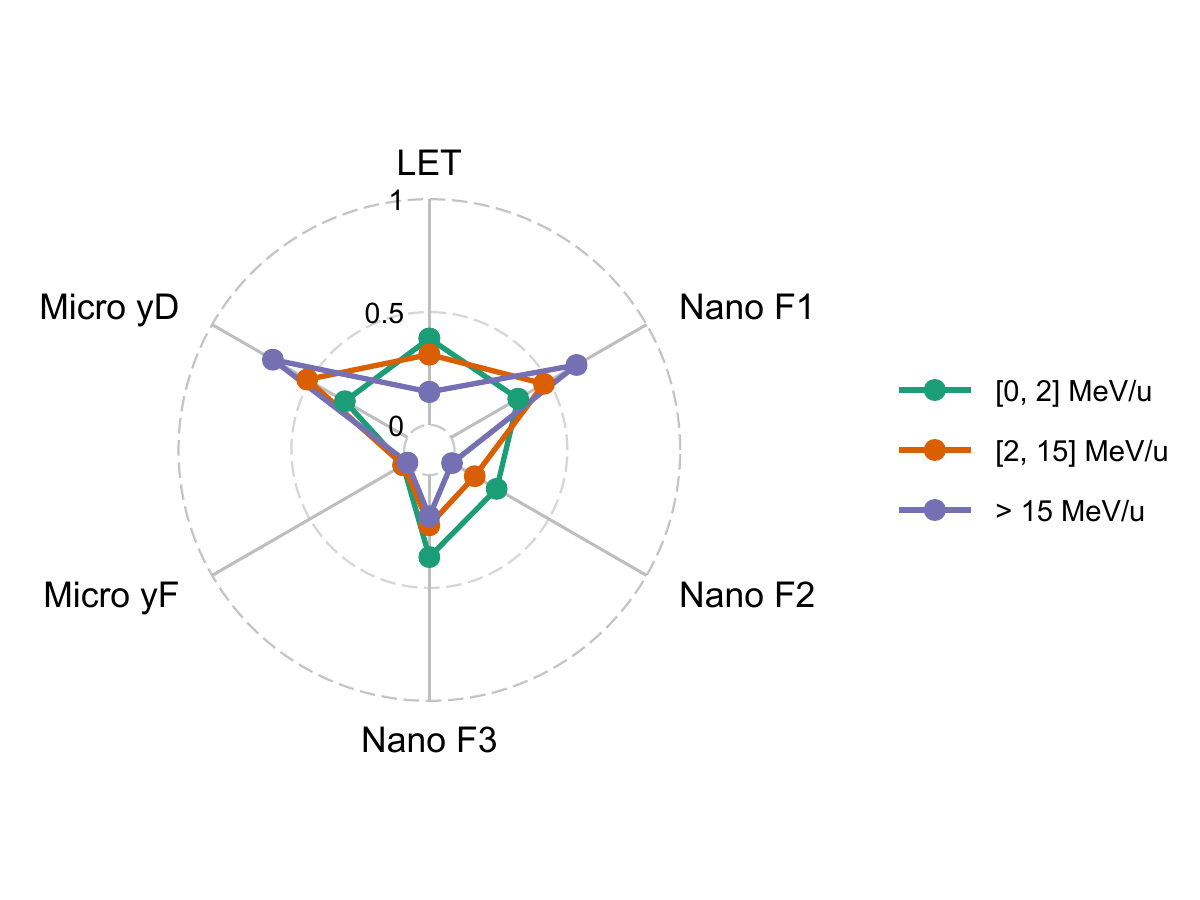}
    \caption{Spider plot with attention for protons at dose corresponding to the low dose limit. Different colors represent different energy bins.}
    \label{fig:RapLDp}
  \end{subfigure}
  \hfill
  \begin{subfigure}[t]{0.48\textwidth}
    \centering
    \includegraphics[page=1,width=\linewidth]{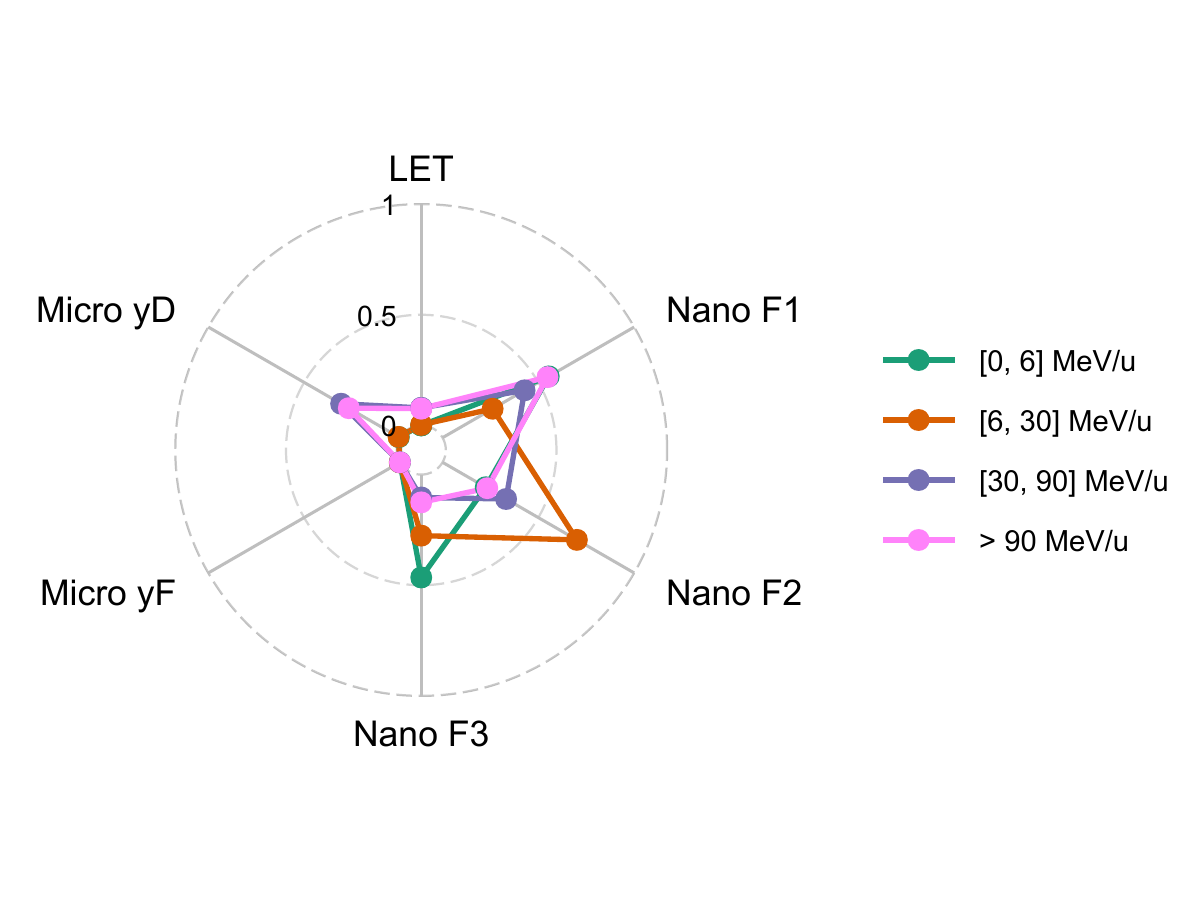}
    \caption{Spider plot with attention for carbon ions at a dose corresponding to the low dose limit. Different colors represent different energy bins.}
    \label{fig:RadLDc}
  \end{subfigure}
  \caption{Spider plot with attention for protons and carbon ions for the three endpoints considered.}
  \label{fig:Rad}
\end{figure}

We further analyzed feature attention using spider plots, grouping experiments by particle energy as indicated in the color legend. Figures~\ref{fig:Rad10p}–\ref{fig:RapLDp} show results for protons at $D_{10}$, 2~Gy, and the low-dose limit, with attention weights normalized to the total for each endpoint. Different colors represent energy bins, illustrating how feature importance varies with beam energy. For protons, microdosimetric contributions increase with energy, driven primarily by $\bar{y}_F$, while $\bar{y}_D$ remains comparable at lower energies for $D_{10}$ and 2~Gy. In the low-dose limit, nanodosimetry, particularly $F^*(1)$, dominates across all energies, with only $\bar{y}_F$ approaching similar importance.

The same analysis for carbon ions (Figures~\ref{fig:Rad10c}–\ref{fig:RadLDc}) confirms the overall dominance of nanodosimetric descriptors over other scales. At higher energies, microdosimetry and LET gain relative importance, whereas at lower energies $F^*(1)$ and $F^*(3)$ remain the most influential for $D_{10}$ and 2~Gy. In the low-dose region, nanodosimetry again prevails, with attention shifting toward $F^*(1)$ as energy increases and decreasing for $F^*(3)$; $F^*(2)$ shows a peak at intermediate energies.

\begin{figure}[htbp]
  \centering
  \begin{subfigure}[t]{0.48\textwidth}
    \centering
    \includegraphics[page=1,width=\linewidth]{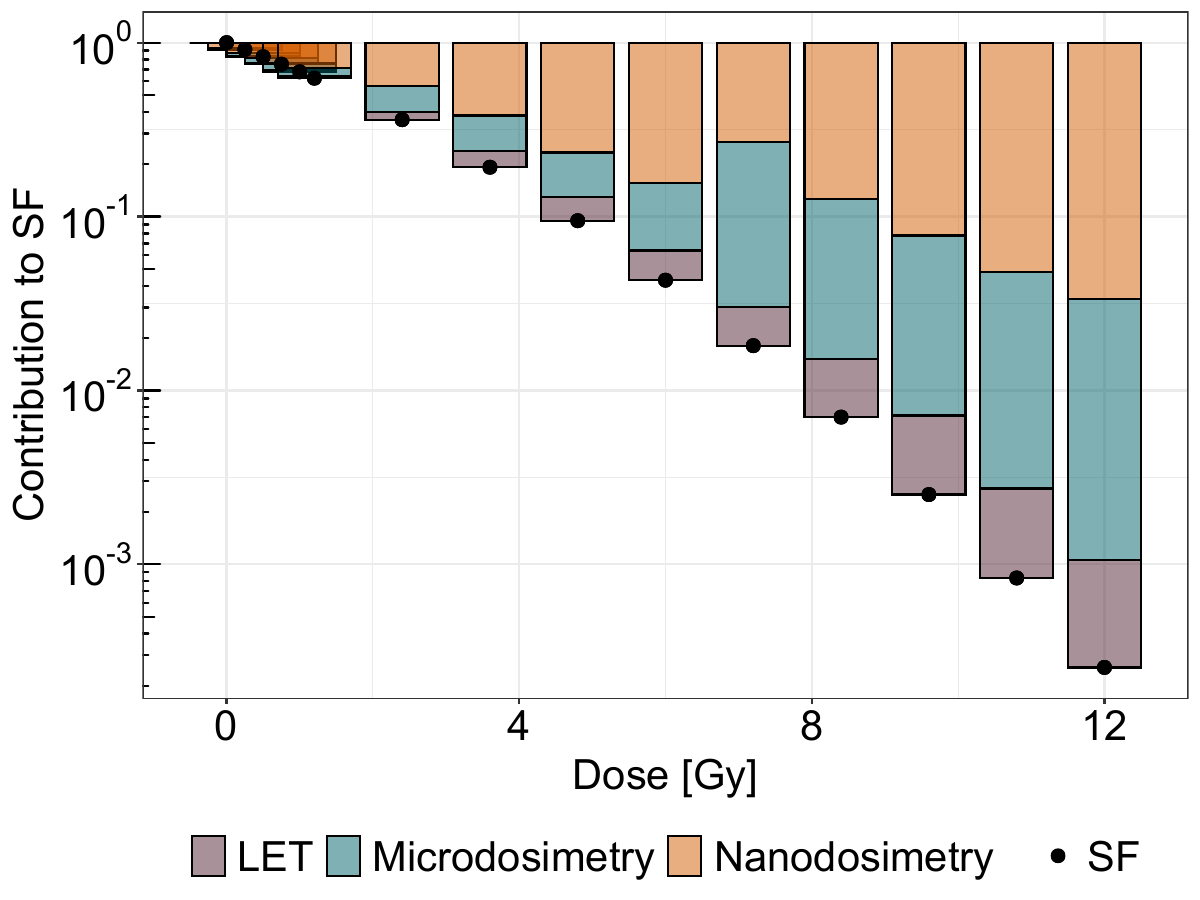}
    \caption{Contribution of the different spatial scales, reported by different colors, to SF calculations for carbon ions irradiation on V79 cell line at LET 22 keV$/ \mu$m.}
    \label{fig:V791}
  \end{subfigure}
  \hfill
  \begin{subfigure}[t]{0.48\textwidth}
    \centering
    \includegraphics[page=1,width=\linewidth]{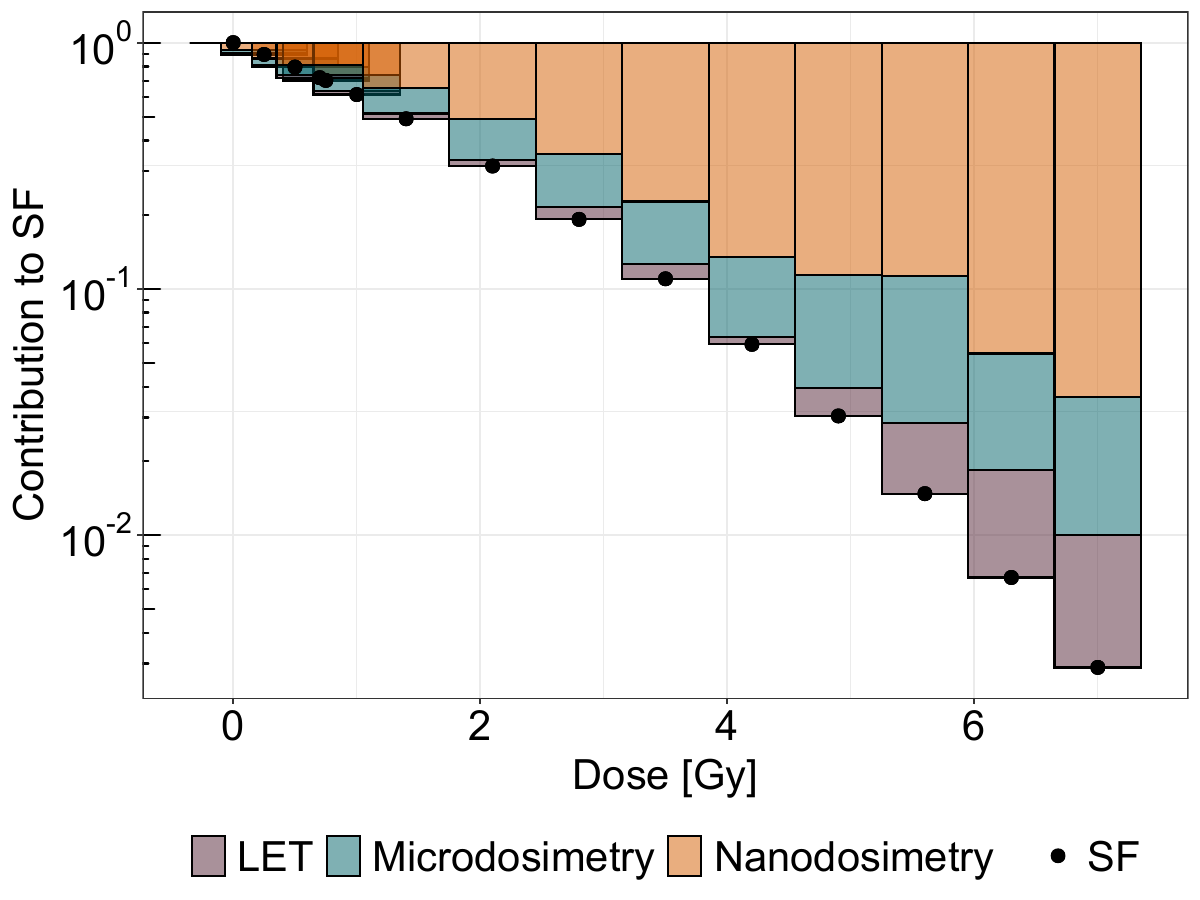}
    \caption{Contribution of the different spatial scales, reported by different colors, to SF calculations for carbon ions irradiation on V79 cell line at LET 41 keV$/ \mu$m.}
    \label{fig:V792}
  \end{subfigure}
  \hfill
  \begin{subfigure}[t]{0.48\textwidth}
    \centering
    \includegraphics[page=1,width=\linewidth]{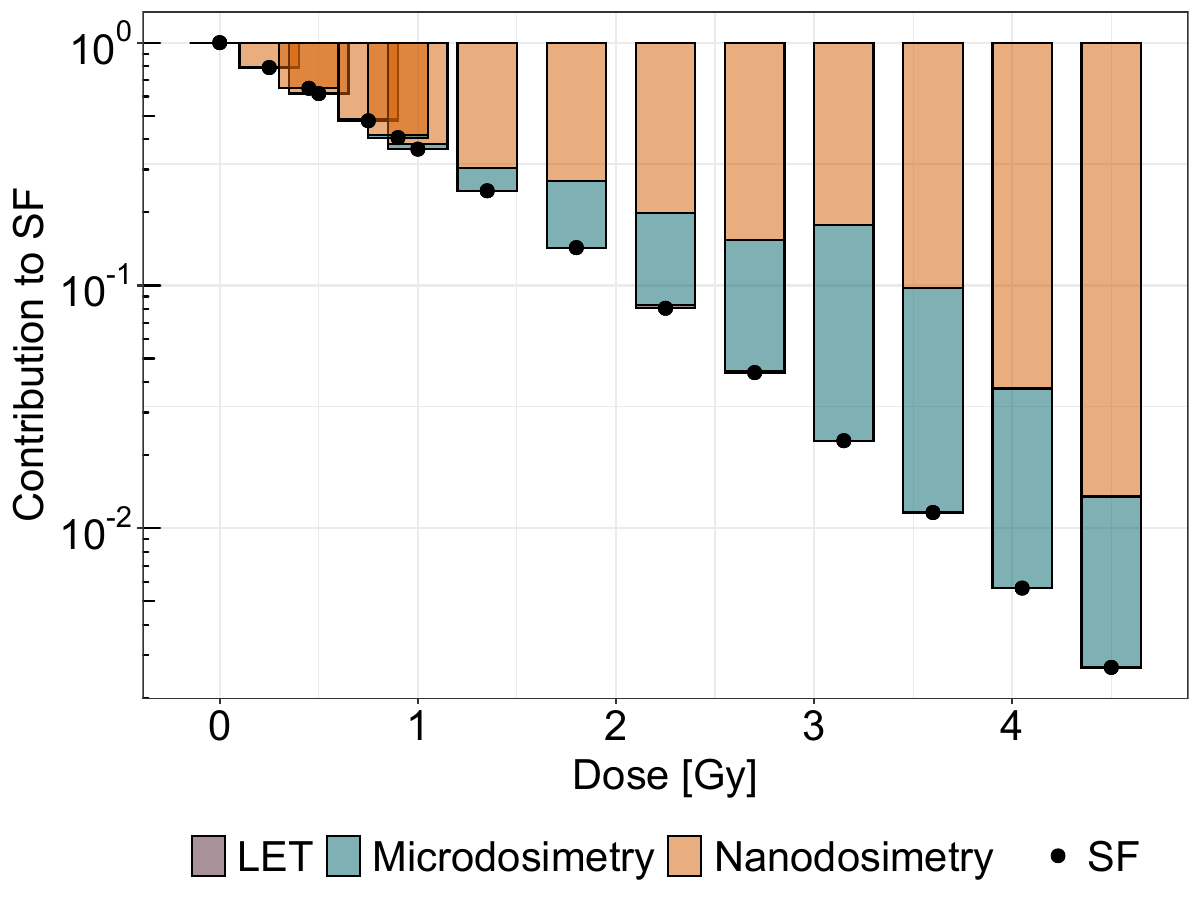}
    \caption{Contribution of the different spatial scales, reported by different colors, to SF calculations for carbon ions irradiation on V79 cell line at LET 142 keV$/ \mu$m.}
    \label{fig:V793}
  \end{subfigure}
  \hfill
  \begin{subfigure}[t]{0.48\textwidth}
    \centering
    \includegraphics[page=1,width=\linewidth]{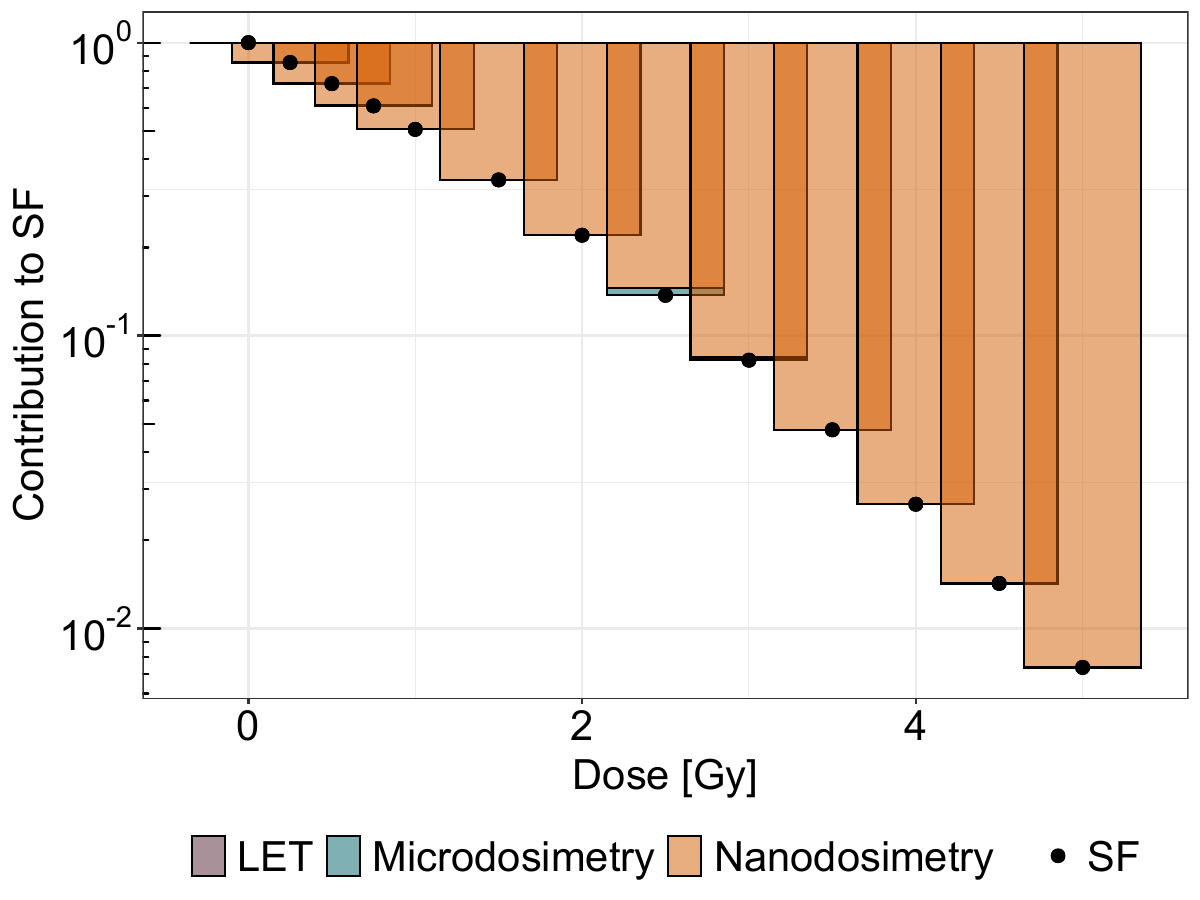}
    \caption{Contribution of the different spatial scales, reported by different colors, to SF calculations for carbon ions irradiation on V79 cell line at LET 360 keV$/ \mu$m.}
    \label{fig:V794}
  \end{subfigure}
  \caption{Contribution of the different spatial scales, reported by different colors, to SF calculations for carbon ions irradiation on V79 cell line at four different LET values.}
  \label{fig:V79}
\end{figure}

\begin{figure}[htbp]
  \centering
  \begin{subfigure}[t]{0.48\textwidth}
    \centering
    \includegraphics[page=1,width=\linewidth]{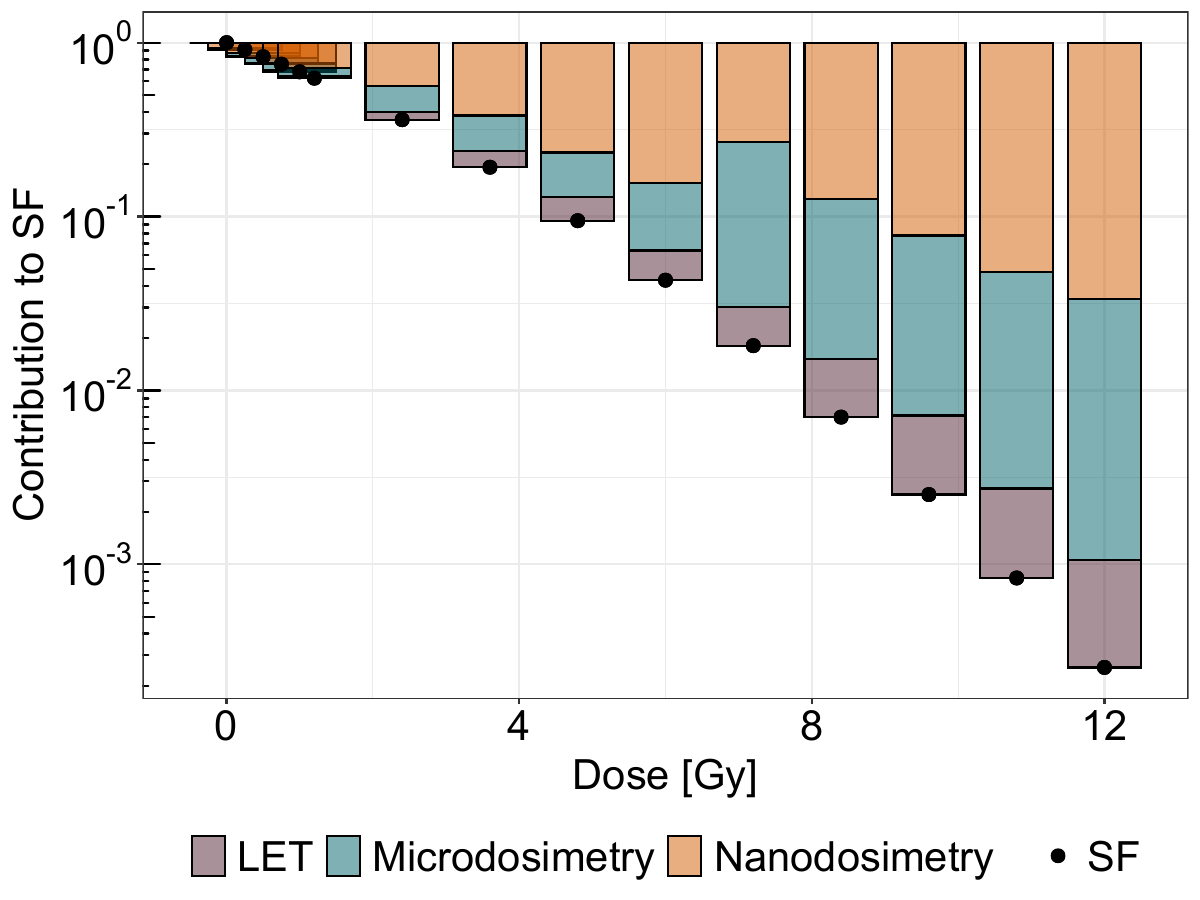}
    \caption{Contribution of the different spatial scales, reported by different colors, to SF calculations for carbon ions irradiation on HSG cell line at LET 23 keV$/ \mu$m.}
    \label{fig:HSG1}
  \end{subfigure}
  \hfill
  \begin{subfigure}[t]{0.48\textwidth}
    \centering
    \includegraphics[page=1,width=\linewidth]{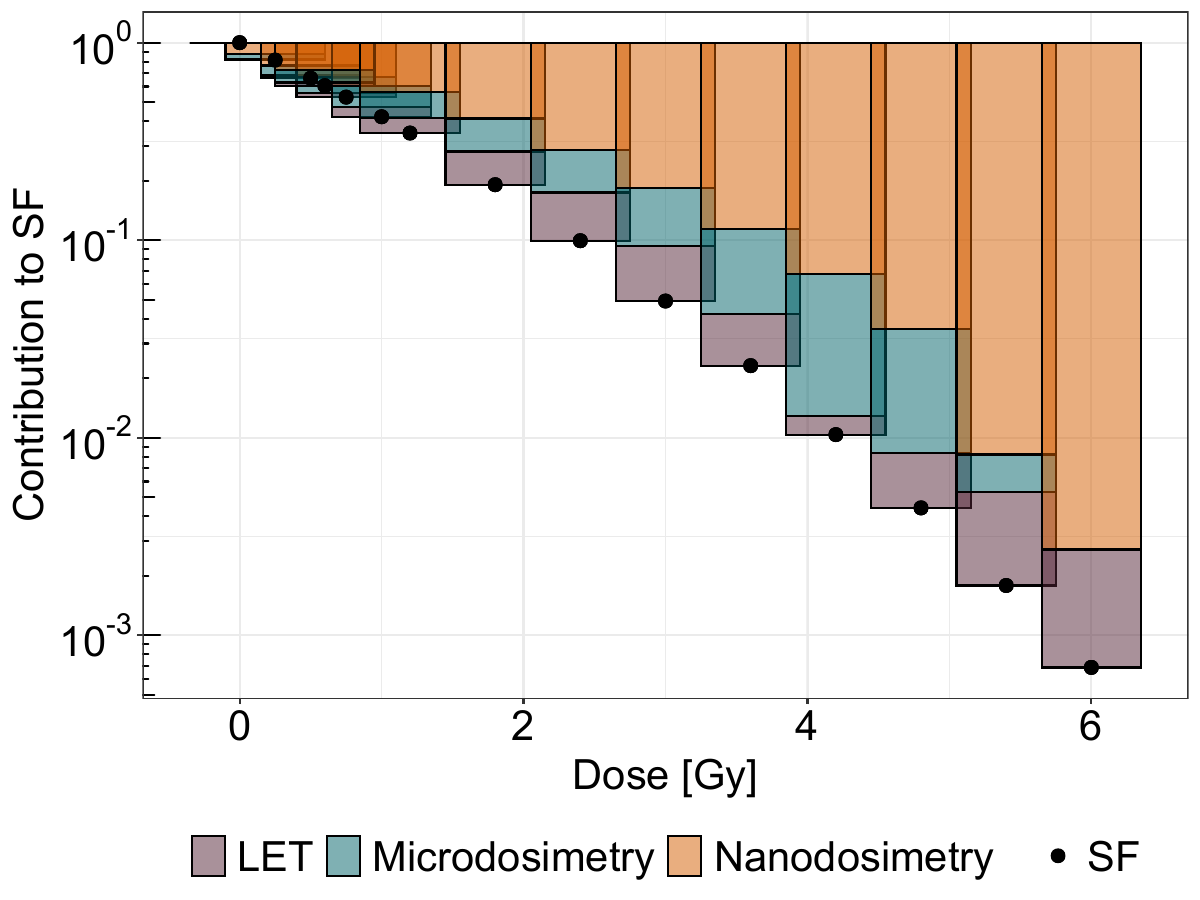}
    \caption{Contribution of the different spatial scales, reported by different colors, to SF calculations for carbon ions irradiation on HSG cell line at LET 40 keV$/ \mu$m.}
    \label{fig:HSG2}
  \end{subfigure}
  \hfill
  \begin{subfigure}[t]{0.48\textwidth}
    \centering
    \includegraphics[page=1,width=\linewidth]{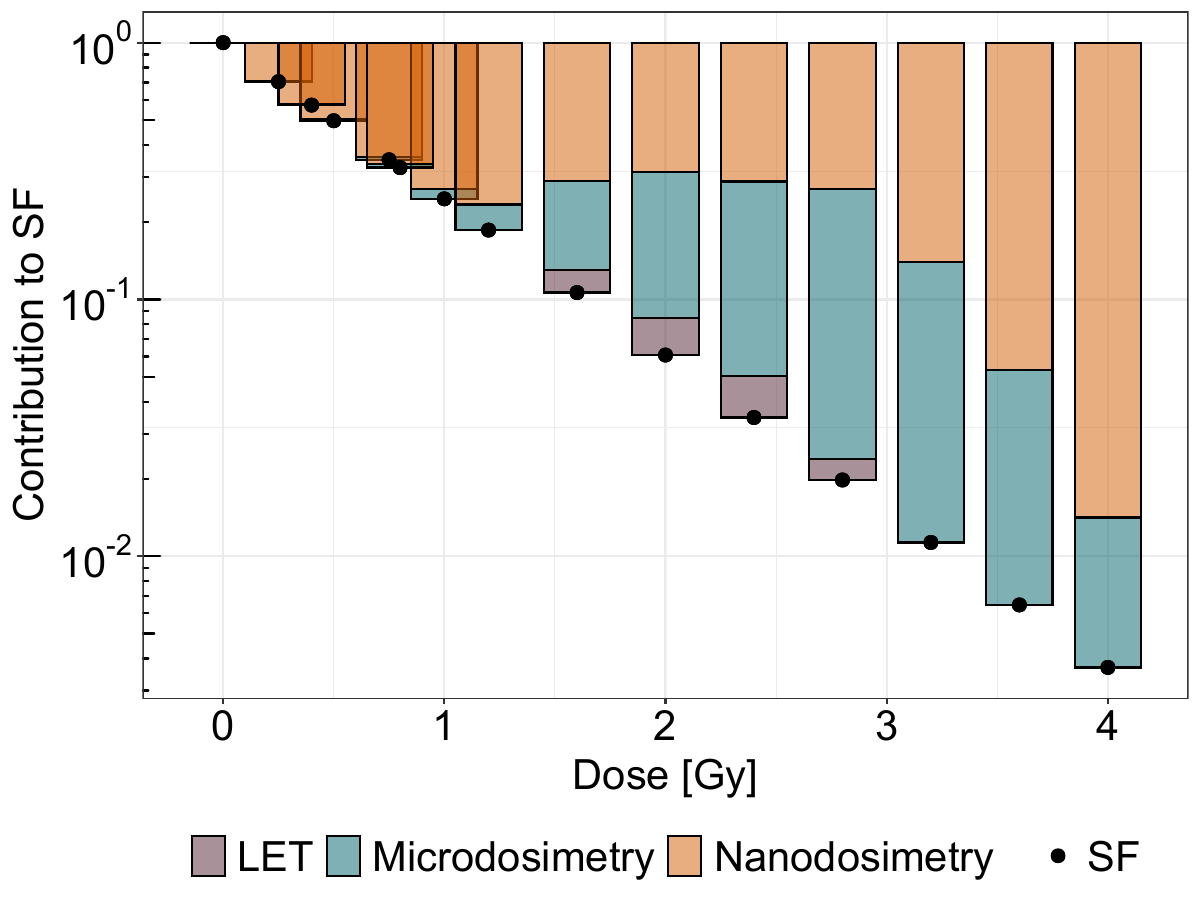}
    \caption{Contribution of the different spatial scales, reported by different colors, to SF calculations for carbon ions irradiation on HSG cell line at LET 144 keV$/ \mu$m.}
    \label{fig:HSG3}
  \end{subfigure}
  \hfill
  \begin{subfigure}[t]{0.48\textwidth}
    \centering
    \includegraphics[page=1,width=\linewidth]{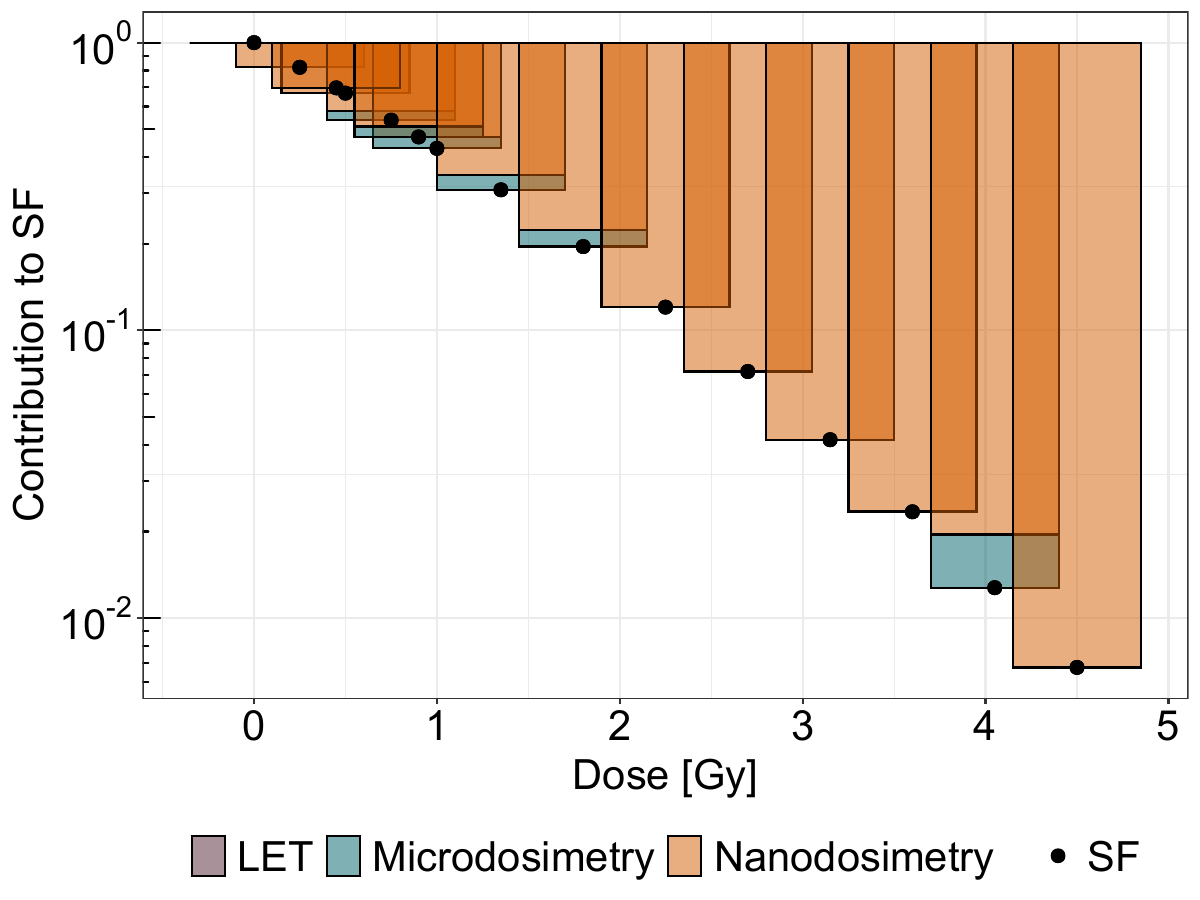}
    \caption{Contribution of the different spatial scales, reported by different colors, to SF calculations for carbon ions irradiation on HSG cell line at LET 467 keV$/ \mu$m.}
    \label{fig:HSG4}
  \end{subfigure}
  \caption{Contribution of the different spatial scales, reported by different colors, to SF calculations for carbon ions irradiation on HSG cell line at four different LET values.}
  \label{fig:HSG}
\end{figure}

Figure \ref{fig:V79} shows the contribution of different spatial scales to the prediction of the surviving fraction values for the V79 cell line. All experiments are taken from \cite{furusawa2000inactivation}. We consider data from one single experiment with carbon ions irradiation corresponding to four different LET values: \ref{fig:V791} 22 keV/$\mu$m, \ref{fig:V792} 41 keV/$\mu$m, \ref{fig:V793} 142 keV/$\mu$m, and \ref{fig:V794} 360 keV/$\mu$m. The predicted SF points are represented with black dots, while colored bars represent the relative contribution of different scales to the overall survival.
We show the same results in Figure \ref{fig:HSG} for the HSG cell line. In this case the LET values are \ref{fig:HSG1} 23 keV/$\mu$m, \ref{fig:HSG2} 40 keV/$\mu$m, \ref{fig:HSG3} 144 keV/$\mu$m and \ref{fig:HSG4} 467 keV/$\mu$m.

\begin{figure}[htbp]
  \centering
  \begin{subfigure}[t]{0.48\textwidth}
    \centering
    \includegraphics[page=1,width=\linewidth]{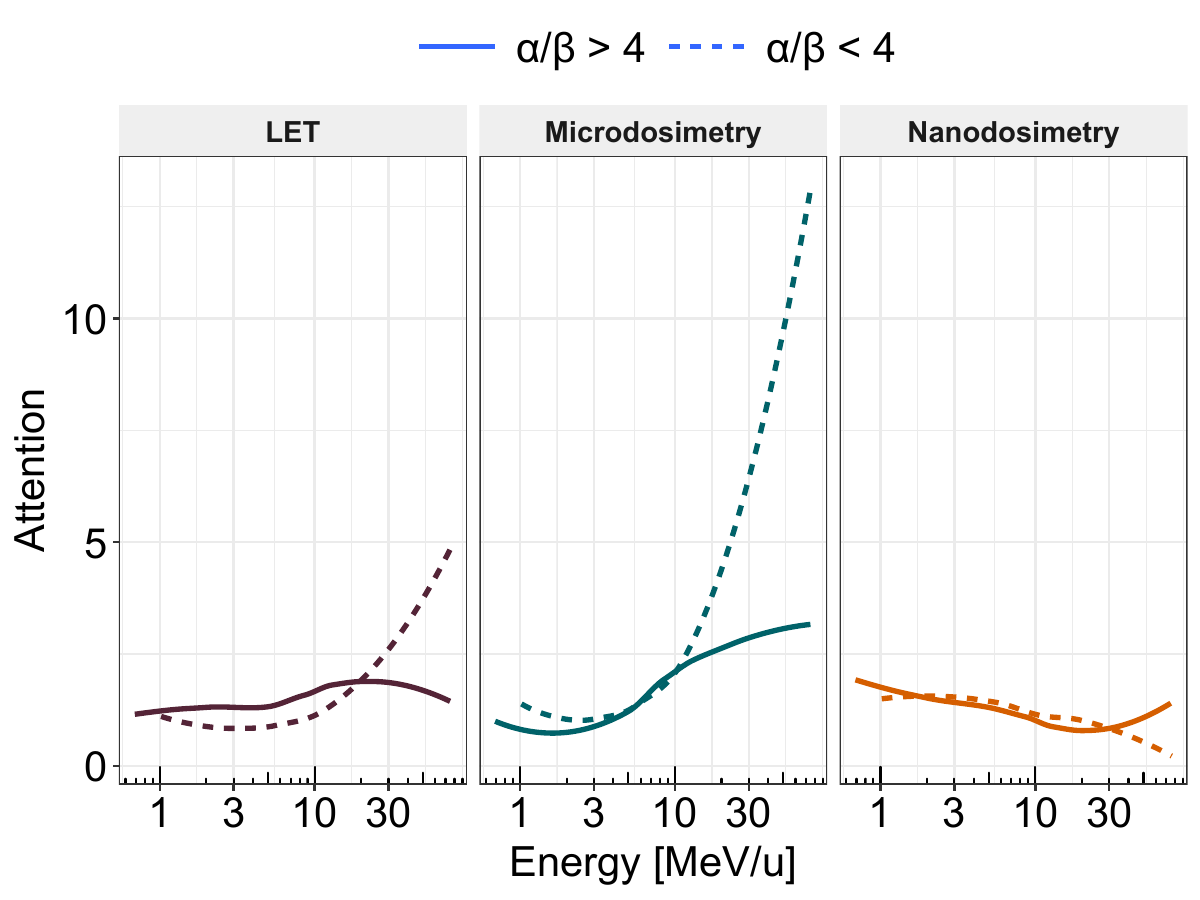}
    \caption{Attention to LET, microdosimetric, and nanodosimetric features evaluated at dose corresponding to 10$\%$ survival as a function of beam energy for protons. Linetype denotes different cell-line radiosensitivity expressed in terms of $\alpha_X/\beta_X$ ratio.}
    \label{fig:EngAttAB10}
  \end{subfigure}
  \hfill
  \begin{subfigure}[t]{0.48\textwidth}
    \centering
    \includegraphics[page=1,width=\linewidth]{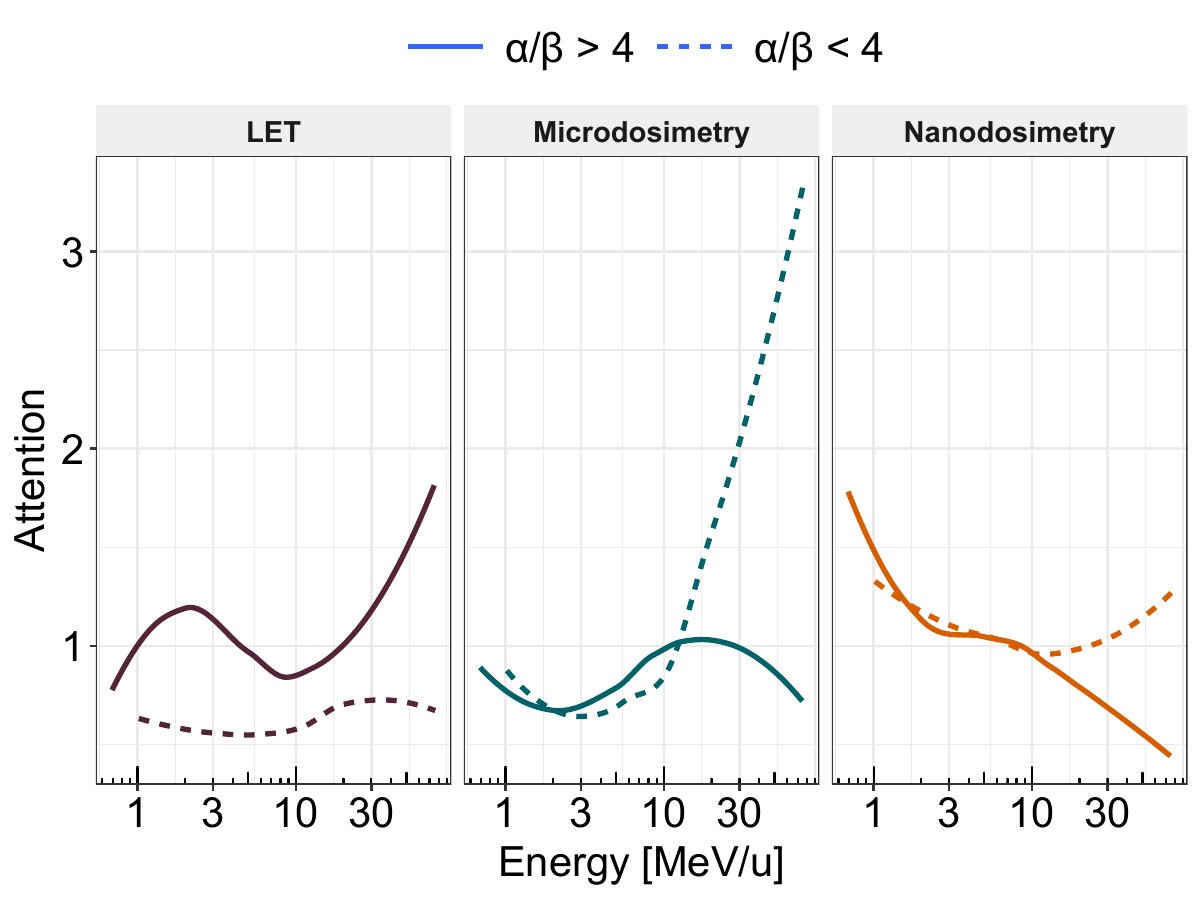}
    \caption{Attention to LET, microdosimetric, and nanodosimetric features evaluated at 2 Gy as a function of beam energy for protons. Linetype denotes different cell-line radiosensitivity expressed in terms of $\alpha_X/\beta_X$ ratio.}
    \label{fig:EngAttAB2}
  \end{subfigure}
  \hfill
  \begin{subfigure}[t]{0.48\textwidth}
    \centering
    \includegraphics[page=1,width=\linewidth]{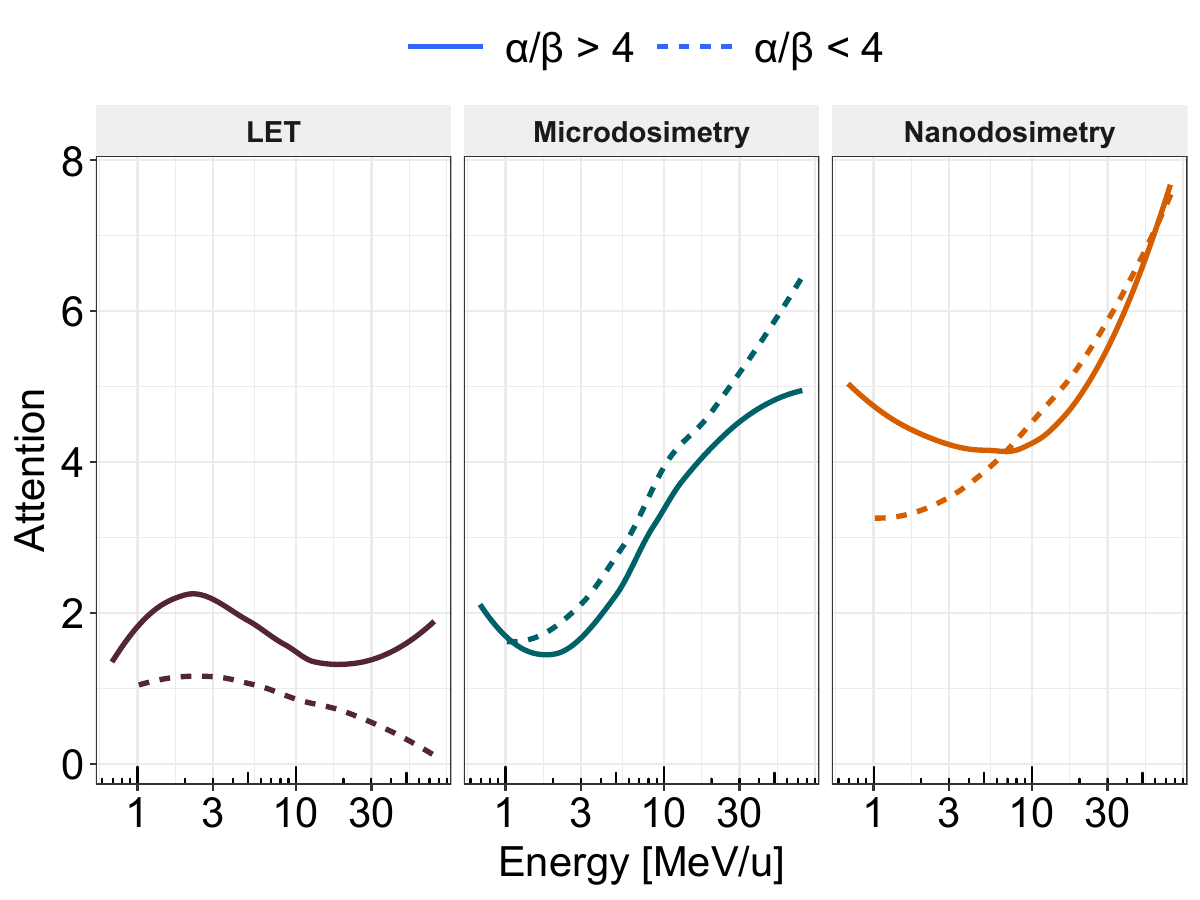}
    \caption{Attention to LET, microdosimetric, and nanodosimetric features evaluated at dose corresponding to the low dose limit as a function of beam energy for protons. Linetype denotes different cell-line radiosensitivity expressed in terms of $\alpha_X/\beta_X$ ratio.}
    \label{fig:EngAttABLD}
  \end{subfigure}
  \caption{Attention as a function of beam energy for protons for the three endpoints considered divided by $\alpha_X/\beta_X$ ratio.}
  \label{fig:EngAttAB}
\end{figure}

Figure~\ref{fig:EngAttAB} illustrates attention as a function of particle energy, separated into panels for LET, microdosimetry, and nanodosimetry for protons at $D_{10}$ (Figure~\ref{fig:EngAttAB10}), 2~Gy (Figure~\ref{fig:EngAttAB2}), and the low-dose limit (Figure~\ref{fig:EngAttABLD}). Line styles distinguish cell lines by radiosensitivity, following the convention in particle therapy: solid lines correspond to $\alpha_X/\beta_X > 4$~Gy, while dashed lines represent $\alpha_X/\beta_X < 4$~Gy. 

For $D_{10}$ and 2~Gy, LET and microdosimetry show relatively stable attention for highly radiosensitive cell lines ($\alpha_X/\beta_X > 4$~Gy), but increase with energy for less sensitive lines ($\alpha_X/\beta_X < 4$~Gy). Nanodosimetry exhibits minimal differences between cell types, with only a slight decrease in attention at high energies for $\alpha_X/\beta_X < 4$~Gy at 2~Gy. 

In the low-dose limit, attention generally rises across all scales, with nanodosimetry remaining dominant. LET shows an almost constant trend, consistently higher for $\alpha_X/\beta_X > 4$~Gy. Microdosimetry follows a similar increasing pattern, while nanodosimetry demonstrates a strong upward trend with energy, with $\alpha_X/\beta_X > 4$~Gy receiving slightly more attention at low energies.

\section{Discussion}

Radiation response in particle therapy spans multiple biological scales, from clustered DNA damage at the nanometer level, through disruption of chromosomal and cellular structures at the micrometer scale, to normal‑tissue and tumor responses at the macroscopic level. Each scale has inspired distinct modeling approaches that are widely used, yet typically applied predominantly using a single scale.

Nanodosimetry characterizes the stochastic nature of energy deposition at nanometer scales by analyzing ionization cluster-size distributions (ICSD) within DNA-sized volumes~\cite{rabus2011nanodosimetry,grosswendt2007descriptors}. Large clusters are strongly associated with complex, often irreparable DNA lesions, making nanodosimetric descriptors such as cumulative probabilities $F^*(n)$ valuable for linking track structure to biological severity. Recent computational and experimental studies have demonstrated how these metrics can inform treatment planning, for example, in spread-out Bragg peak (SOBP) fields for protons and carbon ions~\cite{ramosmendez2021fast,selva2019nanodosimetry,faddegon2023pmb}.

Microdosimetry addresses energy deposition at the micrometer scale, historically providing the bridge between physical dose and biological effect through probabilistic quantities such as lineal energy ($y$)~\cite{rossi1996microdosimetry,icru2016report}. These concepts underpin the MKM, which remains the clinical standard for calculating RBE in carbon ion therapy, particularly in Japan~\cite{kanai1999mkm}. Microdosimetric descriptors like $\bar{y}_F$ and $\bar{y}_D$ are routinely used to characterize radiation quality and optimize treatment plans~\cite{missiaggia2023investigation, inaniwa2010treatment, bellinzona2021linking}.

At the macroscopic level, LET has long served as a practical descriptor of radiation quality. High-LET radiation, such as carbon ions, produces dense ionization tracks and greater biological impact compared to low-LET radiation like protons or photons. LET optimization strategies exploit this property by modulating beam energies and fluence to increase LET in tumor regions while sparing normal tissue, and recent work has proposed incorporating LET-based objectives into inverse planning~\cite{grupp2023letopt,giantsoudi2016letplanning}.

Together, nanodosimetry, microdosimetry, and LET provide complementary perspectives on radiation quality and biological effect. However, these approaches are almost always applied separately, leaving open the challenge of integrating multiscale information into unified predictive frameworks. Addressing this gap is essential for advancing biologically guided particle therapy.

As a general rule, lower-energy particles tend to be more biologically effective than higher-energy particles due to their higher LET. However, for carbon ions, when LET exceeds approximately 150~keV/$\mu$m (corresponding to energies below about 10~MeV/u), the trend reverses: biological effectiveness decreases despite increasing LET. This phenomenon, known as the \emph{overkill effect}, occurs because excessive ionization within a small volume leads to redundant damage that does not further reduce survival \cite{bellinzona2021linking}.

The results demonstrate that, to achieve accurate predictions of radiation-induced biological effects, the model integrates information across multiple spatial scales, rather than relying on a single physical descriptor of the radiation field. The combined contribution from each of these quantities is essential for capturing the full complexity of radiation–matter interactions and subsequent biological responses. This study supports ongoing efforts to incorporate further physical descriptions of radiation quality into treatment-plan optimization. Historically, biological dose optimization in proton therapy has typically assumed a constant RBE of 1.1, while for carbon ions, models like MKM or LEM have been used to guide planning. However, recent work has shifted towards optimizing not just dose but also physical descriptors such as LET, especially in proton therapy, where LET-guided optimization has already been implemented in clinical settings, and in carbon-ion therapy, where LET painting strategies are being explored to enhance biological effectiveness \cite{deng2021critical, reiners2025rbe}. Beyond LET, there is growing interest in integrating nanodosimetric and microdosimetric quantities into optimization frameworks. For example, cluster-dose optimization for protons and helium has been proposed based on nanodosimetric metrics derived from Monte Carlo track-structure simulations \cite{facchiano2025cluster, faddegon2023ionization}, while carbon-ion plans have been optimized using nanodosimetric quantity-weighted dose approaches \cite{yang2024nanodosimetric}. Similarly, microdosimetric parameters have been shown to provide alternative and effective descriptors of radiation quality in clinical planning rather than conventional LET alone \cite{inaniwa2010treatment, magini2025integration}, enabling novel microdosimetry-guided treatment planning strategies. These efforts underscore a movement toward multi-scale, physics-informed optimization that extends beyond conventional biological-dose models. Moreover, it highlights the value of considering multiple spatial scales rather than relying on a single one. More importantly, it emerges a complex pattern that relates different scales, particle type, energy, and cell-line radiosensitivity, corroborating the idea that several factors contribute to the final radiobiological endpoint.

Concerning the accuracy of the proposed multiscale DL model, it shows improved predictive performance compared to the clinical reference model MKM and LEM approaches. Performance, evaluated using RMSE in Figure \ref{fig:RMSE}, consistently results in lower values for our multiscale model across the considered data from V79 and HSG cell lines. These results suggest that the architecture can better capture the underlying multiscale patterns of the data, leading to a more accurate estimation with respect to the reference models.

The comparison between protons and carbon ions highlights both similarities and fundamental differences in their radiobiological behavior. In general, we could say that $\dt$ and dose 2 Gy show similar attentions, whereas a different pattern emerges in the low dose limit. Although both protons and carbon ions exhibit comparable qualitative trends in the importance of predictors from all spatial scales against beam energy, the relative importance of nanodosimetric quantities is much higher in the case of carbon ions, as shown in Figure \ref{fig:EngAtt}. This effect arises from the characteristic energy deposition along tracks of carbon ions, with higher charge and LET. The importance of nanodosimetry reflects the much denser ionization patterns and more complex DNA damage produced by heavier ions. The different importances across energies suggest that different biological mechanisms can contribute differently to cell inactivation for different particles. Further, shaded areas around the LOESS smoothing reflect biological variability. In this study, all this information is condensed in the $\alpha_X$ and $\beta_X$ parameters for cell survival curves from X-ray irradiation. It is interesting to note how LET in carbon ions shows little to no shaded area, meaning that the DL model treated the LET as a biological invariant offset around other physical descriptions. 

A similar observation applies to the low-dose limit, where nanodosimetry clearly dominates. This trend likely reflects the fact that at very low doses, biological effects are driven by the complexity of individual ionization clusters within DNA-sized volumes. As the dose increases—and consequently the number of particle tracks and interactions grows—macroscopic descriptors such as LET and microdosimetric quantities become increasingly relevant, since they capture cumulative effects and spatial correlations at larger scales.

The dependence between cell survival and radiation field description becomes more complex when the dose is considered, as illustrated in Figures \ref{fig:V79} and \ref{fig:HSG}. Specifically, at lower doses, nanodosimetry plays a more prominent role. As the dose increases, the contribution from the other spatial scales becomes more significant. This trend is observed primarily at lower energies, consistent with earlier observations: low-energy carbon ions exhibit a strong dominance of nanodosimetry, with extremely low energies showing exclusive reliance on this scale. This confirms the considerations done in the differences between $\dt$ and the low dose limit.

A further interesting point to discuss, as shown in Figures~\ref{fig:EngAtt} and~\ref{fig:Rad}, is the presence of a maximum in microdosimetric importance around 10~MeV/u (corresponding to LET $\approx 150$~keV$/\mu$m), which is particularly evident for carbon ions. This LET value is precisely the threshold above which the overkill effect occurs, leading to a decrease in biological effectiveness. The observed attention pattern may be attributed to the complexity of describing biological effects in very high-LET scenarios, where the overkill phenomenon becomes significant for heavy ions. In such cases, the model requires a stronger contribution from microdosimetric descriptors to complement nanodosimetric information in predicting the biological outcome. This behavior supports the hypothesis that microdosimetric parameters act as sensitive indicators of radiation quality, especially in high-LET regimes. It is worth emphasizing that an accurate representation of the overkill effect has been achieved without relying on any saturated microdosimetric term, such as $y^*$, which is commonly used in mathematical models to account for saturation. We deliberately avoided including $y^*$ because this quantity incorporates not only physical aspects but also biological considerations. While $y^*$ is widely adopted in the community to model the overkill effect, our data-driven approach naturally captures this phenomenon without introducing explicit correction terms.

Furthermore, Figure~\ref{fig:Rad} provides additional insights into the specific contributions of each spatial scale. For protons, the impact of microdosimetry—particularly dominant at higher energies—is primarily captured by the fluence-averaged lineal energy \yf. As beam energy increases, the model assigns greater importance to \yf, while the relevance of dose-averaged lineal energy and nanodosimetric descriptors decreases. This suggests that the DL model naturally prioritizes \yf\, over LET, likely because microdosimetry offers a more accurate representation of large particle tracks produced by high-energy protons compared to the more aggregated LET metric. In such cases, fluence-averaged quantities provide a more representative characterization than dose-averaged metrics.
In the low-dose limit, an interesting pattern emerges: the importance of $F^*(1)$ increases with energy, whereas $F^*(3)$ exhibits the opposite trend. This behavior is consistent with the physics of track structure—low-energy protons tend to produce larger ionization clusters, while higher-energy protons generate smaller clusters, making single-ionization probabilities more relevant at high energies. Furthermore, the contribution from microdosimetry is primarily driven by $\bar{y}_d$, compared to $\dt$ and the dose at 2~Gy, where the majority of the contribution is associated with $\bar{y}_f$. This observation is consistent with the MKM, which postulates that the $\alpha$ parameter depends on radiation quality through $\bar{y}_d$. Since the $\alpha$ parameter corresponds to the low-dose limit of the logarithmic survival curve, this provides the same information as the attention observed in the low-dose region.

Regarding carbon ions, generally, the main responsibility for nanodosimetry is $F^*(1)$, as evident in Figure \ref{fig:Rad10c}, showing that a first-order quantity is already a good indicator of DNA damage clustering, with an enhanced strong contribution of $F^*(3)$ at extremely lower energies that it is most likely used by the model to predict the biological effect at extremely low energies.

Beyond characterizing the radiation field itself, our results emphasize how variations in radiosensitivity among different cell lines lead to distinct responses under the same irradiation conditions. Radiosensitivity is quantified through the ratio $\alpha_X/\beta_X$ for a reference radiation, as commonly adopted in the community. This ratio, derived from the LQ model, is a key radiobiological parameter: tissues with a high $\alpha_X/\beta_X$ value (e.g., most tumors) are dominated by the linear component and exhibit reduced sensitivity to fractionation, whereas tissues with a low $\alpha_X/\beta_X$ value (e.g., late-responding normal tissues) are more sensitive to dose per fraction. This concept is fundamental in treatment planning, guiding fractionation strategies to balance tumor control and normal tissue sparing. While a threshold of approximately 10~Gy is typically considered, our dataset contains few experiments with $\alpha_X/\beta_X > 10$~Gy. Therefore, we adopt a threshold of 4~Gy for classification purposes. The analysis is restricted to protons because the cell lines used for carbon ions in the dataset are not uniformly tested across all energies, introducing bias. It is important to note that, in general, cell lines with higher $\alpha_X/\beta_X$ ratios are less sensitive to variations in particle energy and, consequently, to differences in particle efficacy. This observation aligns with the current understanding that particles with higher LET are more robust against variations in cell-line radiosensitivity. Furthermore, attention at the nanodosimetric scale appears to be largely independent of the cell line, whereas the most significant differences for $\Delta t$ and 2~Gy emerge at the microdosimetric level for higher-energy protons. In the low-dose limit, there is an increasing emphasis across the three scales—from LET to microdosimetry and nanodosimetry. Notably, LET plays a more critical role for cell lines with high $\alpha_X/\beta_X$, whereas microdosimetry becomes more relevant for those with low $\alpha_X/\beta_X$.

In general, this work has been inspired by the long-standing effort of the community to relate energy deposition to biological effect, as well as by recent results reported in \cite{friedrich2018dna}, where the authors pursued a similar goal through a model-based analysis. Our work shares the same overarching objective but adopts a purely data-driven perspective. Some analogies and differences emerge between the two approaches. 

In particular, regarding contributions in the low-dose limit, \cite{friedrich2018dna} reports a dominance of the micrometer scale for high-energy carbon ions and of the nanometer scale for low-energy carbon ions. We observe similar trends in that nanodosimetry dominates for low-energy carbon ions, with reduced absolute attention at higher energies. However, even at high energies, nanodosimetry remains the most influential descriptor in our model. 

Concerning the more macroscopic descriptor, which in \cite{friedrich2018dna} corresponds to a measure around 10~$\mu$m, whereas in our case it is LET, we consistently report lower attention. Several possible explanations for this discrepancy exist, but we believe that the different characterization of the radiation field is the most relevant. Specifically, we employ microdosimetric and nanodosimetric descriptors, whereas \cite{friedrich2018dna} relies on the amorphous track model \cite{friedrich2013local}, which parameterizes radial energy deposition. This choice reflects the methodological foundations: the authors in \cite{friedrich2018dna} build upon the LEM model, while our data-driven approach offers greater flexibility to incorporate diverse predictors.

One of the key strengths of our approach lies in the use of the attention mechanism to offer native interpretability through its sequential attention-based feature selection mechanism. Unlike post-hoc explainability methods such as SHAP, which often require building surrogate models and rely on permutation-based techniques, our approach provides transparent insights directly from the model itself. This is particularly advantageous in our context, where the correlation of different radiation quality metrics undermines the reliability of permutation-based explanations. In our model, feature correlations are addressed intrinsically through the sparsemax-based attention mechanism implemented in TabNet. This enforces sparsity in the attention maps, compelling the model to focus on a limited subset of features at each decision step, before a subsequent aggregation with different weights. This allows the model to capture diverse and complementary information while mitigating the effects of feature collinearity. At each of the three steps in the model developed in this research, a sparsemax activation function is applied to enforce selective attention to a limited number of variables. This mechanism selects features and mitigates issues related to feature correlation. Each step produces a prediction, and these predictions are subsequently recombined into the final output. Details on the model are given in the Appendix.
This built-in transparency not only enhances trust in the model’s predictions but also facilitates its potential translation into clinical settings, where interpretability is essential for decision support and regulatory acceptance.

One limitation of this work is that, although we extracted the beam type (monoenergetic or SOBP) from the PIDE database, this classification is inherently constrained. In practice, so-called monoenergetic beams are often produced through passive energy degradation, meaning they do not represent truly pristine beams. As a result, the distinction between monoenergetic and SOBP beams may be blurred, introducing ambiguity in beam characterization. Moreover, perfect beam characterization is not possible when relying on historical datasets like PIDE, which often lack detailed metadata about beam delivery and modulation.

Another important consideration is the biological endpoint. Although the chosen endpoint represents the current gold standard, it is not necessarily the most relevant from a radiobiological perspective. Other endpoints, closer to in vivo conditions, could provide more clinically meaningful insights and should be integrated when data become available. The main drawback here is data availability; however, our approach is designed to be extensible and could incorporate additional endpoints if suitable datasets exist.

At last, it is worth mentioning that the investigation presented in this work relies on a set of radiation-field metrics that we believe are among the most relevant for describing the physical determinants of biological response. However, other descriptors could certainly be considered. Our choice was motivated by the limited size of the available dataset, which constrains the number of features that can be reliably included without introducing overfitting. To address this, we restricted the feature set while maintaining a balance between microdosimetric and nanodosimetric representations. In general, alternative selections are possible, and future studies with larger datasets could explore a broader range of physical descriptors to further enhance model performance and interpretability.

Finally, transparency is fundamental to ensure clinical translation. This is another strength of our approach: by maintaining interpretability and openness in the modeling process, we facilitate reproducibility and foster trust in the integration of AI into particle therapy.

Overall, our findings advocate for an integrative framework that unites physics-based descriptors at different spatial scales with biological cell characterization, ultimately advancing the precision and personalization of particle therapy treatments. To fully realize this potential, we also emphasize the need for more comprehensive and standardized reporting of radiobiological measurements. Improved documentation would enable more accurate Monte Carlo-based characterization of experimental conditions, particularly regarding beam properties and microenvironmental factors. Given our observation that multiple spatial scales significantly influence cell survival outcomes, it becomes increasingly clear that capturing all relevant scales consistently across experiments is challenging, yet essential for robust and interpretable modeling.

\section{Conclusions}

This study demonstrates that attention-based DL can accurately predict cell survival and RBE in hadron therapy by integrating radiation descriptors across nanodosimetric, microdosimetric, and macroscopic scales. 
Attention analysis shows that no single scale dominates; instead, the model dynamically combines complementary information, with nanodosimetric and microdosimetric quantities being most influential and LET contributing modestly. 
The ion-specific attention patterns confirm the need for a multiscale approach to capture both microscopic and macroscopic effects. Our framework achieves high predictive accuracy (MAPE $\approx$ 10$\%$) while integrating interpretability, highlighting AI as a valuable tool for radiobiological modeling and future clinical translation.

The present study highlights the potential of a data-driven approach to unravel the complex interplay between different scales of energy deposition, dose, and biological endpoints. These interactions reflect the underlying mechanisms of DNA damage induction, repair, and cell population dynamics that ultimately shape cell survival. Similar, if not more intricate, mechanisms can be expected for clinically relevant endpoints. Artificial intelligence offers a transformative opportunity to incorporate such detailed physical and biological descriptions into treatment-plan optimization. Importantly, this work demonstrates that AI is not only capable of achieving high predictive accuracy, as widely recognized, but can also provide transparency and interpretability, enabling us to understand the model’s reasoning. This capability is fundamental for clinical adoption and trust in AI-driven decision support.

%\appendix
%\section*{Appendix A}
%
%\subsection*{Model Hyper-parameters}
%
%\begin{itemize}
%    \item \textbf{Number of attention steps:} Typical values are between 3 and 10. More steps allow the model to extract complex feature interactions but %may lead to overfitting or longer training times.
%    \item \textbf{Learning rate:} Controls the step size during optimization. Common range: $0.001$ to $0.1$.
%    \item \textbf{Sparsity regularizer:} Encourages attention over a small subset of features. Typical values: $10^{-5}$ to $10^{-2}$.
%\end{itemize}
%\subsection*{Training Parameters}

\cleardoublepage
%\nocite{*}
\bibliographystyle{apalike}

\bibliography{references}

\end{document}